\documentclass[12pt,preprint]{aastex}
\begin{document}
\title{Planetary nebulae in the elliptical galaxy NGC 821: 
    kinematics and distance determination\thanks{Based on
    data collected at the Subaru Telescope, which is 
    operated by the National Astronomical Observatory of Japan.}}

\author{A. M. Teodorescu\altaffilmark{1}, R. H. M\'endez\altaffilmark{1},
        F. Bernardi\altaffilmark{2}, A. Riffeser\altaffilmark{3}, 
        and R. P. Kudritzki\altaffilmark{1}}

%   \offprints{R.H. M\'endez}
\email{ana@ifa.hawaii.edu, mendez@ifa.hawaii.edu}

\altaffiltext{1}{Based on data collected at the Subaru Telescope, which is 
operated by the National Astronomical Observatory of Japan.}

\altaffiltext{1}{Institute for Astronomy,
      University of Hawaii, 2680 Woodlawn Drive, Honolulu, HI 96822, USA}

\altaffiltext{2}{Universit\'a di Pisa, Largo B. Pontecorvo 5, 56127,
                 Pisa, Italy}

\altaffiltext{3}{Universit\"ats-Sternwarte M\"unchen, Scheinerstr. 1, 
      D-81679 M\"unchen, Germany}

%   \date{Received ... ; accepted ...}

\begin{abstract}
%Version 12 August 2010.

Using a slitless spectroscopy method with the 8.2 m Subaru telescope 
and its FOCAS Cassegrain spectrograph, we have increased 
the number of planetary nebula (PN) detections and PN velocity 
measurements in the flattened elliptical galaxy NGC 821. 
A comparison with the detections reported previously by the 
Planetary Nebula Spectrograph group indicates that we have confirmed 
most of their detections. The velocities measured by the two groups, 
using different telescopes, spectrographs and slitless techniques, 
are in good agreement. 
We have built a combined sample of 167 PNs and have confirmed the 
keplerian decline of the line-of-sight velocity dispersion
reported previously. We also confirm misaligned rotation
from the combined sample.
A dark matter halo may exist around this galaxy, but 
it is not needed to keep the PN velocities below the local escape 
velocity as calculated from the visible mass.
We have measured the $m$(5007) magnitudes of 145 PNs and produced a 
statistically complete sample of 40 PNs in NGC 821. The resulting PN
luminosity function (PNLF) was used to estimate a distance modulus of 
31.4 mag, equivalent to 19 Mpc. We also estimated the PN formation rate.
NGC 821 becomes the most 
distant galaxy with a PNLF distance determination. The PNLF distance 
modulus is smaller than the surface brightness fluctuation (SBF) 
distance modulus by 0.4 mag. Our 
kinematic information permits to rule out the idea that a shorter 
PNLF distance could be produced by the contamination of the PNLF 
by background galaxies with emission lines redshifted into the 
on-band filter transmission curve. 
\end{abstract}

   
\section{INTRODUCTION}

Planetary nebulae (PNs) are easy to detect in the outskirts of galaxies 
less distant than 25 Mpc. 
Once detected, the strong emission lines in PN spectra are well
suited for accurate radial velocity measurements. Thus, PNs are useful
test particles to study dark matter existence and distribution in 
halos of elliptical galaxies (Hui et al. 1995; M\'endez et al. 2001;
Romanowsky et al. 2003; Teodorescu et al. 2005; De Lorenzi et al. 
2008, 2009; M\'endez et al. 2009; Coccato et al. 2009).

Photometry of the detected PNs permits to build their luminosity 
function (PNLF), which can be used to provide a distance estimate. 
PNLF distances play a significant role as a critical link 
between the Population I and Population II distance scales 
(Ciardullo et al. 2002; Feldmeier et al. 2007). One unsolved problem
involving the PNLF is that Cepheid-calibrated PNLF distances are 0.3 mag
smaller than Cepheid-calibrated surface brightness fluctuation (SBF)
distances (Ciardullo et al. 2002). We would like to understand why.

PNs in the elliptical galaxy NGC 821 were first studied by 
Romanowsky et al. (2003), and most recently by Coccato et al. (2009). 
Using the Planetary Nebulae Spectrograph (PN.S)
at the La Palma 4.2 m Herschel telescope, they discovered 127 PNs 
which they subsequently used to study the kinematics of this galaxy. 

The results of the dynamical studies have not been quite conclusive 
about the presence of dark matter in normal (intermediate mass; not 
giant) elliptical galaxies. Some of the normal ellipticals show clear 
evidence of a dark matter halo, like NGC 5128 (Hui et al. 1995; 
Peng et al. 2004). In some other cases, the line-of-sight velocity 
dispersion (LOSVD) shows a keplerian decline as a function of 
distance from the center, which can indicate either the absence 
of a dark matter halo, or the presence of radial anisotropy in
the velocity distribution. Examples are NGC 4697 (M\'endez et al.
2001, 2009), NGC 3379 (De Lorenzi et al. 2009), and NGC 821. 
The degeneracy between mass and radial anisotropy has not been 
resolved to everybody's satisfaction. Dekel et al. (2005) say that 
radial anisotropy is expected from 
their numerical simulations, while Kormendy et al. (2009), in their
comprehensive study of the Virgo cluster ellipticals, say that the
normal ellipticals (like the ones that have presented the Keplerian 
decline of LOSVD) are more isotropic than giant ellipticals (which
do frequently show dark matter halos).
There is general agreement that radial anisotropy, if present, can 
explain the Keplerian decline of the LOSVD; but we believe it is fair 
to say that nobody has produced decisive independent observational 
evidence of radial anisotropy, so that the issue remains undecided
(De Lorenzi et al. 2009; Napolitano et al. 2009).

In the particular case of NGC 821 there is apparently conflicting 
evidence from a long-slit spectroscopic study by Forestell \& Gebhardt 
(2008), which seems to indicate a non-decreasing LOSVD with radius, 
in contradiction to what the PNs suggest. Unfortunately, the 
Forestell \& Gebhardt data do not extend beyond 100 arcsec from the center,
so it is not possible to compare with the LOSVD from PNs at more than 200
arcsec from the center (Coccato et al. 2009).
More recently, two additional absorption-line studies involving 
NGC 821 have been published. (1) Proctor et al. (2009) used data extracted 
from Keck DEIMOS multi-object spectroscopy, extending out to about 120
arcsec along the major axis. (2) Weijmans et al. (2009) used the 
integral-field spectrograph SAURON to extend a previous SAURON study
by Emsellem et al. (2004). The data from Weijmans et al. also stop
at 120 arcsec along the major axis, and 50 arcsec along the minor axis 
of NGC 821. Both groups concluded that their LOSVD is in good general 
agreement with Forestell \& Gebhardt, leaving aside a few discrepant 
points, and they also found good agreement with PN kinematics within 
their accessible range.

Proctor et al. (2009) did not attempt to discuss the dark matter halo. 
Weijmans et al. (2009) built Schwarzschild dynamical models and concluded
that their best-fit model requires a dark matter halo. Unfortunately, 
they chose to ignore the PN data points, for consistency, 
they said, because 
the PN measurements are not based on integrated stellar light. So
their Figure 14 stops at 120 arcsec from the center of NGC 821, and we 
cannot see how well their best-fit model is able to fit the substantially 
lower PN LOSVD at 200 arcsec. 
As a consequence, we are left with the impression that the kinematic 
information provided by PNs is still irreplaceable, but on the other 
hand it is still being rejected sometimes as unreliable. 

Our original motivation was to make a deeper PN survey in NGC 821, and 
use the expected increase in the number of PN detections to improve the 
statistics and reach a clearer conclusion about the dark matter content
in this particular case. After two observing runs with the 8.2 m Subaru 
telescope and its Cassegrain imaging spectrograph FOCAS, we were able
to increase the number of detected PN candidates, but not as much as 
we hoped, because of bad weather and poor seeing in the second run.
We concluded that a sophisticated modeling effort would again fail to 
give a decisive contribution. Instead, we focused on providing some 
useful information.
 
Our goals in this paper are: (1) to compare our detections, 
PN positions and radial velocities with those measured by the PN.S 
group with a different telescope, spectrograph and technique; which
might contribute to an easier acceptance of the PN kinematic results.
(2) to use our photometry to build for the first time the PNLF of 
NGC 821, estimate its PNLF distance, and compare it with the SBF 
distance. Since recently many SBF distances in the Virgo and 
Fornax clusters were redetermined using the Hubble Space Telescope 
Advanced Camera for Surveys (Blakeslee et al. 2009), we think it is 
timely to revisit the PNLF--SBF distance comparison.

Section 2 describes our observations, reductions and radial velocity
calibrations. In Section 3 we compare our results with those obtained by 
the PN.S group. Sections 4, 5 and 6 present the kinematic results from 
167 PNs and their interpretation. In Section 7 we describe the 
PN photometry. In Section 8 we build the PNLF, make a PNLF 
distance determination and estimate the PN formation rate. 
Section 9 deals with the persistent discrepancy between PNLF and SBF 
distances. In Section 10 we give a summary of our conclusions.

\section{OBSERVATIONS, REDUCTIONS, AND RADIAL VELOCITY CALIBRATIONS}

\subsection{Observations}

The observations were made by one of us (RHM) with the Faint Object
Camera and Spectrograph (FOCAS; Kashikawa et al. 2002) attached 
to the Cassegrain focus of the 
8.2 m Subaru telescope, Mauna Kea, Hawaii, on three nights, 2004 November 
7, 8 and 9. The nights were dark, of photometric quality, and with average 
seeing of 0$''$.7, 1$''$, and 0$''$.5 on the 
first, second and third night, respectively.

The field of view of FOCAS is 6.5 arcmin and is covered by two CCDs of 2k  
$\times$ 4k (pixel size 15  $\mu$m) with an unexposed gap of 5$''$ 
between them. For simplicity, we will call the 2 CCDs Chip 1 and Chip 2.
The image scale is 0.104 arcsec pixel$^{-1}$.
The purpose of this project was to discover the PNs using the classical 
on-band, off-band filter technique (see Section 2.2);
then measure the brightness and the radial
velocities of the discovered sources. The radial velocities were measured 
using a slitless method (see Section 2.2) involving images taken 
through both the on-band filter and an echelle grism.
The on-band filter has a central wavelength of 5025 \AA, a FWHM of 60 \AA, 
a peak transmission of 0.68 and an equivalent width of 40 \AA.
In total, 15 off-band images (exposure time 140 s), 16 on-band images 
(exposure time 1400 s) and 16 grism + on-band images (exposure time 2100s) 
were taken. For the photometric
calibration, on-band images of the spectrophotometric standard LTT 9491 
(Oke 1990) were taken. For the wavelength calibration, after 
each grism exposure, an engineering mask
was inserted in the light path, and on-band and grism + on-band images 
were obtained illuminating the mask with the comparison lamp. 
In addition, on-band and grism+on-band images of the engineering mask
were taken illuminating the mask with
the Galactic PN NGC 7293 (PNG 036.1-57.1), for radial velocity
quality control. Examples of the calibration images are shown in Figure 2 
of M\'endez et al. (2009). Table 1 shows the log of the most important 
images used in this work.

\subsection{PN detection and slitless spectroscopy with Subaru and FOCAS}

The traditional on-band, off-band filter technique was used for the detection
of the PNs in NGC 821. The on-band image is taken through a narrow-band 
filter passing the redshifted [O III] $\lambda$5007 nebular emission line, 
while the off-band image is taken through a broader filter passing no nebular 
emissions. The PNs are visible as point sources in the on-band image, but 
are absent in the off-band image. A third image, taken through the on-band 
filter and a grism, confirms the PN candidates. By inserting the grism in 
the light path, the images of all continuum sources are transformed into 
segments of width determined by the on-band filter transmission curve. All 
the emission-line point sources such as the PNs remain as point sources. The 
grism also introduces a shift relative to the undispersed on-band image which 
is a function of the wavelength of the nebular emission line and of position 
on the CCD. By calibrating this shift, we are able to measure the radial 
velocities for all emission-line objects in the field. In the case of Subaru 
and FOCAS the dispersing element was an echelle grism with 175 grooves/mm 
which operates in the 4th order and gives a dispersion of 0.5 \AA/pixel, 
with an efficiency of 60\% (see Subaru FOCAS website). Typical images with 
seeing around 0.5 arc second have a Point Spread Function size of 
5 pixels, which translates into a radial velocity resolution of 140 
km s$^{-1}$, {\it i.e.} the PN's internal velocity field is not resolved. 
Assuming position errors of 0.4 pixel, the expected uncertainty in radial 
velocity is 12 km s$^{-1}$.

\subsection{Data reduction}

Standard IRAF\footnote{IRAF is
distributed by the National Optical Astronomical Observatories,
operated by the Association of Universities for Research in
Astronomy, Inc., under contract to the National Science
Foundation} tasks were used for the basic CCD reductions 
(bias subtraction, flat-field correction using twilight flats).
In order to eliminate the cosmic rays and to detect faint PN candidates, 
we needed to combine the scientific images of NGC 821: First, for each field,
Chip 1 and Chip 2, we chose as reference images one pair of undispersed and
dispersed on-band individual images of the best possible quality. All the 
other available images, including the off-band ones, were registered on the
corresponding reference image. Any possible displacements due to guiding 
problems or deformations in the spectrograph were thus reduced to a minimum.
The registration was done with the IRAF tasks ``geomap'' and ``gregister'' 
and about 30 stars for each field were used. In the end, we obtained 
residuals smaller than 0.2 pixel.

The registration of the dispersed images was more difficult. Since we could
not detect enough PNs on the individual images, we had to instead use the 
spectral segments produced by normal stars for the registration. To minimize
the temperature shifts in the position of the spectral segments, we divided
the grism images into groups having approximately the same temperature
(which was obtained for each grism image from the corresponding FITS header). 
This time, the registration proceeded in two steps: First, the images in each 
group were registered. Then, the resulting registered images of each group 
were registered on the image with the best quality.    
We calculated the centroid of each segment using the FORTRAN 90 program 
``centroid'' written by David Tholen.
About 10 stars were used for both fields and residuals around 0.3 pixel were
obtained. An error of 0.4 pixel in the distance between undispersed and 
dispersed images would produce an error of 12 km s$^{-1}$ in radial 
velocity. Having obtained a satisfactory registration, we produced the 
combined on-band, off-band and dispersed images for Chip 1 and Chip 2 fields
using the IRAF task ``imcombine''.

For easier PN detection and photometry in the central parts of NGC 821, where 
the background varies strongly across the field, we produced difference images 
between undispersed on-band and off-band combined frames. In ideal conditions 
this image subtraction should produce a flat noise frame with the 
emission-line sources as the only visible features. A critical requirement to 
achieve the desired result is perfect matching of the point-spread functions
(PSFs) of the two frames to be subtracted. For this purpose, we applied a 
method for ``optimal image subtraction'' developed by Alard \& Lupton 
(1998), and implemented in Munich by G\"ossl \& Riffeser (2002) as part of 
their image reduction pipeline. Figure 1 shows part of the resulting 
difference image from Chip 2. 

This procedure cannot be used for the combined dispersed images because 
there is no off-band counterpart. Therefore, to flatten the background and 
reduce the contamination by stellar spectra, we applied the 
IRAF task ``fmedian'' to the combined Chip 1 and Chip 2 dispersed images.
The resulting median images were then subtracted from the unmedianed ones. 
Figure 2 shows the result for the same field shown in Figure 1. 

\subsection{Astrometry}

We identified the PN candidates by blinking the on-band versus the off-band 
difference images and confirmed them by blinking on-band versus dispersed. In 
addition, the object had to be a point source and to be undetectable in the 
off-band image. In this way it is possible to minimize the contamination of 
the PN sample by unrelated background sources, like galaxies with emission 
lines redshifted into the on-band filter transmission curve. We will 
further discuss background contamination in Sections 5 and 8. The pixel 
coordinates of all the candidates in the undispersed 
and dispersed images were measured with the IRAF task ``phot'' with the 
centering algorithm ``centroid''.

We performed an astrometric calibration of the images using the
USNO-B1 astrometric star catalog (Monet et al. 2003). 
We found 11 astrometric stars on
Chip 1 and 13 astrometric stars on Chip 2. We used a software written
by David Tholen which is able to perform up to a quartic fit of the
stars in a field. This software is usually used to produce very accurate
astrometry needed for determining the positions of asteroids.  Because
of the relatively small size of the field in a chip, a simple linear
fit was sufficient.  The astrometric rms of the fit is composed by
three components: (1) the contribution of the centroiding error, which
our software is capable to keep under a few tenths of a pixel (a few
hundredths 
of an arcsecond) even for objects with a S/N near 5; (2) the contribution 
of the random error of the stars, which, for a single star is of the
order of 0$''$.2, but which is going with the square root of the
number of stars used (in our case the total contribution is of the
order of few hundredths of an arcsecond); (3) the systematic error of the 
catalog, which is estimated to be around 0$''$.2. 
We estimate an rms of about 0$''$.3 for the astrometry of our catalog 
of PNs.

\subsection{Radial velocity calibrations}

In order to determine the shift produced by the insertion of the grism
as a function of wavelength and position on the CCD, we used an engineering
mask that produces an array of point sources when it is illuminated with
the internal FOCAS lamps or any extended astronomical source. The full mask
is made up of almost 1000 calibration points, separated by almost 100 pixels.
We also used exposures of NGC 7293, a local PN with a large 
angular size that allowed us to obtain calibration measurements all across 
the FOCAS field. An example of these calibration images is shown in Figure 2 
of M\'endez et al. (2009). The procedure for wavelength measurement is 
explained in Section 3 of that paper, to which we refer the interested 
reader.

To test for the presence of any systematic errors in the radial 
velocities, we use the images of NGC 7293 as follows. We can 
measure radial velocities in two ways: 
(a) classical, using each mask hole as a slit, and (b) slitless, 
using the displacement as a measure of wavelength and therefore velocity.
The comparison between slitless versus classical measurement is shown
in Figure 3 for both Chip 1 and Chip 2. Since we are measuring the 
velocity of different gas elements in NGC 7293, we expect to get a 
range of velocities across the field. 
There is good agreement between classical and slitless measurements. 
The average heliocentric velocity of NGC 7293 from all the grid points 
is about $-$30 km s$^{-1}$, in good agreement with recent measurements of
the systemic velocity ($-$27 km s$^{-1}$; Meaburn et al. 2005). For a more 
detailed discussion, please see M\'endez et al. (2009). We conservatively 
estimate the calibration errors in FOCAS slitless radial velocities to be 
of the order of 10 km s$^{-1}$. 

If we add quadratically the uncertainties in velocity given by the 
calibration errors ($\sim$ 10 km s$^{-1}$, the position errors 
($\sim$ 10 km s$^{-1}$, Section 2.2), and the errors from image registration 
($\sim$ 10 km s$^{-1}$, Section 2.3), we get an overall error of about 
17 km s$^{-1}$. 
Assuming that the spectrograph deformations and guiding errors have a 
marginal contribution, we estimate the total uncertainty in the velocities 
measured with Subaru to be at most 20 km s$^{-1}$.

\section{THE FOCAS SAMPLE AND COMPARISON WITH THE PN.S SAMPLE}

We have found 155 PN candidates in NGC 821 with Subaru and FOCAS. The list 
of detected objects is given in Table 2. We identified 85 PNs in common with 
the PN.S sample, and those are indicated in Table 2. We could not confirm a 
detection for 19 PN.S objects within the FOCAS field we observed.
There are 23 PN.S objects reported outside of the FOCAS field or in places
we could not see (the 5$''$ gap between Chips 1 and 2, and some bad 
columns, mostly in Chip 1). 
Thus we have been able to increase the total number 
of PN candidates from 127 reported by the PN.S group to 178.
Figure 4 shows the distribution of the PN candidates across NGC 821. In  
Figure 5 we present the comparison of equatorial coordinates measured by
both groups, and in Figure 6 we compare the velocities.
In Figure 5 we observe an offset of almost 2$''$ in $\alpha$ between 
the PN.S sample and the FOCAS sample, while the differences in $\delta$ 
give reasonably good agreement within 1$''$. 

The offset and higher residuals in $\alpha$ between the PN.S 
and the FOCAS sample are most probably caused 
by uncertainty in the zero point of the PN.S astrometry. In the PN.S 
counterdispersed imaging method there is no undispersed image, and
the position of a reference star is determined by calculating the 
centroid
of the stellar spectrum segments in the counterdispersed images. 
Since the stars are dispersed in the E-W direction, it becomes more 
difficult to measure the Right Ascension. For example, the centroid
calculation may be sensitive to the presence of absorption features
near one edge of the filter transmission curve, and this will depend
on the spectral types of the stars used (L. Coccato 2009, private 
communication).

In Figure 5 we have indicated 8 cases (identified with the FOCAS numbers 
in Table 2) where we have detections lying close to reported PN.S 
detections. However, since $\Delta\alpha$ and $\Delta\delta$ are 
too large, we do not consider these PN.S detections to be confirmed 
by our images. There is one marginal case, our object 242, which
lies at the edge of the elongated cloud of confirmed sources. However,
in this case our velocity is completely different, as shown in Figure 6;
because of that, we do not count this as a confirmation of the PN.S
detection. In fact, this PN.S object (number 92 in their list) 
was considered an outlier by the PN.S team, and was not included 
in the final sample they used for their kinematic study.

Figure 6 shows that there is good agreement between both sets of 
velocities. Having eliminated object 242 from the comparison, 
we find a standard deviation of 31 km s$^{-1}$, which is expected,
given the error bars reported by both groups.

\section{PRELIMINARY ANALYSIS OF PN RADIAL VELOCITIES}

We will refer to our heliocentric radial velocities determined with the 
slitless method simply as 'velocities'. We could measure the velocities 
for 140 of the 155 PN candidates listed in Table 2 (some objects could 
not be measured on the grism images because of background contamination 
problems, or bad columns, or because the displacement produced by the 
grism put the sources outside of our field). Since we 
found good agreement between the velocities measured with 
both techniques, as shown in Figure 6, we decided to use the PN.S 
velocities in those few cases (five) where a detection was confirmed but
we could not make a velocity 
measurement. We also added the 23 PN.S velocities measured outside 
of the FOCAS field or in places we could not see.

After these additions,
the total sample of PN candidates (FOCAS + PN.S) discovered in NGC 821
with measured velocities amounts to 168. Figure 7 shows the ($x$,$y$) 
coordinates of the 168 PN candidates relative to the 
center of NGC 821, adopted to be at $\alpha (2000) = $ 2h 08m 21.4s,
$\delta (2000) = 10^\circ 59' 41''$. The $x$ coordinate runs in the 
direction of increasing $\alpha$ along the major axis of NGC 821, defined 
to be at PA = 25$^\circ$ (from N through E).
Figures 8 and 9 show the PN candidate velocities as a function of the 
$x$-coordinate in arcsecs and the $y$-coordinate in arcsecs, respectively.
The average velocity of 1706 km s$^{-1}$ is in good agreement, within 
the uncertainties, with the NASA/IPAC Extragalactic Database 
(NED) radial velocity of 1735 km s$^{-1}$. 
We estimate the PN average velocity uncertainty to be approximately 20
km s$^{-1}$, from a velocity dispersion of the order of 150 km s$^{-1}$
(see Section 5), the number of PNs measured, and the possible systematic 
error of $\pm$ 10 km s$^{-1}$ in our velocities from the calibration 
procedure using NGC 7293.

In Figures 8 and 9, we find one object with a too low velocity if 
interpreted as a 5007 \AA \ emitter: it is FOCAS number 
104 in Table 2. It may appear to be a marginal case, but we will see in 
next section, in the plot related to escape velocities, that this object
has the wrong velocity for NGC 821, and it must be a background 
galaxy with some emission line that has been redshifted into the on-band 
filter transmission curve. Therefore we decided to reject object 104
and work with a sample of 167 PNs with known velocities.

\section{LINE-OF-SIGHT VELOCITY DISPERSIONS AND ESCAPE VELOCITIES}

As already mentioned, Romanowsky et al. (2003) and Coccato et al. (2009)
found a decreasing line-of-sight velocity dispersion (LOSVD) 
which could imply a 
dearth of dark matter in this galaxy. We want to verify if our velocities 
confirm this result. For that purpose we subdivide our 167 PN 
sample into five elliptical annuli, with shapes similar to that of 
NGC 821, at increasing angular distances from the center of NGC 821.
The numbers of PNs per annulus, from the inside out, are 48, 48, 23, 27,
and 19, respectively.
For all PNs within each elliptical annulus we calculate the average 
angular distance to the center, and the LOSVD. 
The result of this calculation is shown in Figure 10. It looks
very similar to plots shown by Romanowsky et al. (2003) and Coccato et al. 
(2009), so essentially we have confirmed their results. Figure 10 also
shows LOSVDs derived from long-slit absorption-line spectra of
NGC 821 (Forestell \& Gebhardt 2008). We also added SAURON data 
taken from Weijmans et al. (2009). We find marginal agreement, within 
error bars, between PNs and absorption-line data within 100$''$ of the 
galaxy's center. 
We have tried several different versions of Figure 10, using different 
sets of PN data: vertical regions instead of ellipses, or FOCAS data only,
but the PNs always give more or less the same result, confirming the 
keplerian decline of the LOSVD as previously reported. 

In Figure 10, we fit the run of the LOSVD with angular distance from
the center of NGC 821 using an analytical model developed by Hernquist 
(1990). This model is spherical, nonrotating, isotropic, and it assumes a 
constant mass-to-light ratio (no dark matter). The fit is 
obtained by adopting a total mass of 
2 $\times 10^{11}$ $M_{\odot}$ and an effective radius $R_{e}$ = 39$''$
(Blakeslee et al. 2001), which is equivalent to 4.16 kpc for a distance of 
22 Mpc. Of course this Hernquist model is just a rough first approximation, 
but it has been shown, in another case of a similarly flattened elliptical 
galaxy (NGC 4697, M\'endez et al. 2009), to give a result which is similar, 
within 20\%, to that of the more sophisticated NMAGIC model of De Lorenzi 
et al. (2008). 
%We find no reason to modify the description by Coccato et al. 
%(2009): it may not be possible to rule out the presence of {\it some} dark
%matter around NGC 821, but it will not be as much as predicted by 
%cosmological simulations.

The mass of 2 $\times 10^{11}$ $M_{\odot}$ is 
obtained directly by fitting the LOSVD with the Hernquist model. 
Note that a similar (although slightly larger) mass can be obtained in 
another way: taking the $M/L$ ratio from Gebhardt et al. (2003), as 
tabulated by Napolitano et al. (2005), which is 8.4 in the $B$ band. 
If we combine this with a $B$ luminosity
of 2.7 $\times 10^{10}$ $L_{\odot}$ (which in turn is obtained 
from $M_B$ = -20.6 for NGC 821 and $M_B$ = 5.48 for the Sun), 
we obtain a mass of 2.3 $\times 10^{11}$ $M_{\odot}$.

We can make another test, by plotting PN radial velocities as a function of
angular distance from the center of NGC 821. Suppose we compare with the 
local escape velocity for the Hernquist model with constant mass-to-light 
ratio used in Figure 10. The escape velocity is given by: 

\begin{equation}
V_{\rm esc} = (2 G M_{\rm t} / (r + a))^{0.5}, 
\end{equation}

\noindent where $M_{\rm t}$ is the total mass, and $a$ is a scale length 
equal to $R_{\rm e}/1.8153$. In the presence of a substantial dark matter 
halo, we would expect some PNs to show velocities in excess of the 
escape velocity (which was calculated under the assumption of 
no dark matter), as
it happened in the case of NGC 5128 (Hui et al. 1995; Peng et al. 2004).

The comparison is shown in Figure 11. Even if we reduce the galaxy's mass 
by 20\%, going in the same direction as with NGC 4697 (M\'endez et al. 2009),
we find no object exceeding or even approaching the local escape velocity.
The only exception is FOCAS object 104 in Table 2, 
which is so distant from the 
galaxy's distribution that it must be a background galaxy, as argued above.
Although this kind of argument cannot {\it prove} the absence of dark matter,
it does indicate that any existing dark matter halo is even less conspicuous 
than that of NGC 4697, the evidence for which was in itself inconclusive.
Therefore, it remains quite possible that less massive ellipticals have, 
on average, lower dark matter fractions than the more massive ellipticals
(e.g., Napolitano et al. 2005, 2009), NGC 821 being another example of this 
trend.

\section{ROTATION}

We have used our enlarged PN sample to verify a statement by 
Coccato et al. (2009): that the velocity field of NGC 821 as defined
by the PNs shows a rotation of about 120 km s$^{-1}$ with a misalignment
of 56 degrees between the photometric major axis and the PN kinematic 
major axis. This result differs from the rotation as
indicated by stellar absorption-line data (SAURON team, Emsellem et al. 
2004; and Proctor et al. 2009). The two different data sets are not in 
conflict, however, because the rotational signal from the stars is limited 
to an area within 70 arcsec from the center of NGC 821 (see Figure 18 in 
Proctor et al. (2009), while the rotational signal from the PNs becomes
significant only outside of about 70 arcsec from the center. This is
of course because of the higher LOSVD close to the center, and the 
comparatively small 
number of PNs detected there. But still it is interesting to find the
outer regions showing rotation in a direction that differs from that of 
the central region, and so it is worth checking with a larger PN sample.

The result of the test is shown in Figures 12 and 13. Instead of producing 
a smoothed PN two-dimensional velocity field, as in Coccato et al. (2009), 
we prefer to avoid any smoothing operations, and work with thick cuts
in different directions, which we find easy to interpret. In Figure 12
we have adopted the same orientation as in Figure 3 of Coccato et al.
(2009), with the major axis of NGC 821 in the vertical direction. So the
$x$ and $y$ axes defined for our Figures 7-9 run vertically and 
horizontally, respectively. We selected this orientation because it 
helps to compare our results with those of Coccato et al. The left panel 
of Figure 12 shows a kinematic axis inclined 0 degrees with respect to 
the photometric major axis. We select all PNs within 40 arcsec of the 
kinematic axis, divide them into four bins, and calculate the average 
velocity for each bin, which is then plotted in the right panel. 
The result is no significant rotation in this direction. 

Figure 13 shows two similar plots, corresponding to a kinematic axis
inclined 56 degrees with respect to the photometric major axis. In this
case the four bins produce a significant velocity gradient, as seen in
the right panel. Note how the squares lie on the approaching side and the 
crosses lie on the receding side, exactly as in Figure 3 of Coccato et al.
We have built similar figures for kinematic axes in all orientations 
relative to the photometric major axis, which we do not show for brevity.
We conclude that the maximum rotational signal occurs for an angle of 
56$\pm$20 degrees, as reported by Coccato et al. So we have essentially 
confirmed their results.

\section{PHOTOMETRY}

We could obtain photometry for 145 out of the 155 sources listed in 
our catalog (Table 2).
The [O III] 5007\AA \ fluxes measured through the on-band filter are 
traditionally expressed in 
magnitudes $m$(5007), using the definition introduced by Jacoby (1989),

\begin{equation}
m(5007)=-2.5 \ {\rm log} I(5007) - 13.74. \label{mdef}
\end{equation}

For the flux calibration, we adopted the standard star LTT 9491 (Oke 1990). 
This star has a monochromatic flux at 5025 \AA \ of 1.075 $\times$ 10$^{-14}$ 
ergs cm$^{-2}$ s$^{-1}$ \AA$^{-1}$ (Colina \& Bohlin 1994). 
The flux measured through the on-band 
filter, in units of ergs cm$^{-2}$ s$^{-1}$, can be calculated knowing the 
equivalent width of the on-band filter; using equation (2), we find 
$m$(5007)=17.18 for LTT 9491.

Most PNs were measurable only on the differences of the combined images 
(on $-$ off). Thus, to calculate the $m$(5007) of the PNs we had to go through 
several steps. First, we made aperture photometry of LTT 9491 using the IRAF 
task ``phot''. The FWHM of LTT 9491 was between 3 and 4.5 pixels. We adopted 
an aperture radius of 20 pixels; the sky annulus had an inner radius of 25 
pixels and a width of 5 pixels. The same parameters were used to make aperture 
photometry of four moderately bright stars in the reference images 
corresponding to both fields. These four ``internal standards'' were selected 
relatively distant from the center of NGC 821 to avoid background problems. 

Having tied the spectrophotometric standard to the internal frame standards, 
we switched to strictly differential photometry. We made aperture 
photometry of the internal standards on the Chip 1 and Chip 2 on-band 
combined images to correct for any differences relative to the reference 
images. On the same on-band 
combined images we subsequentely made PSF-fitting DAOPHOT photometry 
(Stetson 1987; IRAF tasks ``phot'', ``psf'' and ``allstar'') of the internal 
standards and four bright PNs. From the aperture photometry and PSF-fitting 
photometry of the internal standards we determined the aperture correction. 
Finally, we made PSF-fitting photometry of all PN candidates on the difference 
images (onband $-$ offband), where the stars had been eliminated. The four 
bright 
PNs were used to tie this photometry to that of the standards. The internal 
errors in the photometry of the difference images were estimated to be below 
5\%. Applying a final correction related to the peak of the on-band filter 
transmission curve (see Jacoby et al. 1987), we obtained $m$(5007) for 
the 145 PNs.

\section{THE PNLF, DISTANCE, AND PN FORMATION RATE}

Having measured the apparent magnitudes $m$(5007), we needed to produce a 
statistically complete sample, because the detectability of a PN varies 
with the background brightness. For this purpose we used a procedure 
already described in Section 5 of M\'endez et al. (2001). In summary,
we eliminated all PNs fainter than $m$(5007)=28.0, beyond which the 
number of PNs per bin shows a marked decrease, and we also eliminated
all PNs within a zone of exclusion characterized by high background counts
and more difficult detectability. This zone of exclusion was an ellipse 
at the center of NGC 821, with minor and major semiaxes of 30 and 
55 arcsec respectively. In this way, we got a statistically complete 
sample of 40 PNs. 

The PN luminosity function (PNLF) was built, using 0.2 mag bins, 
and compared with simulated PNLFs like those used by 
M\'endez \& Soffner (1997) to fit the observed PNLF of M 31. The comparison
is shown in Figure 14. The absolute magnitudes $M$(5007) that produce
the best fit to the simulated PNLF were calculated using 
an extinction correction of 0.385 mag at 5007 \AA \ (from data listed 
in NED; see Schlegel et al. 1998) and a distance modulus 
$m - M$ = 31.4, which is equivalent to 19 Mpc. 
The simulated PNLFs plotted in Figure 14 are binned, like the observed one, 
into 0.2 mag intervals and have maximum final mass of 0.63 $M_{\odot}$, 
$\mu_{\rm max}$ = 1, and sample sizes between 1200 and 
3400 PNs (see M\'endez \& Soffner 1997; the ``sample size'' is the total 
number of PNs, detected or not, that exist in the surveyed area).
We estimated an error of 0.1 mag from the goodness of the fit at 
different distance moduli. To obtain the total error estimate, we have to 
combine the possible systematic and random errors. The 
systematic error is the same as in Jacoby et al. (1990), i.e., 0.13 mag, 
including the possible error in the distance to M 31, in the modeling of the 
PNLF and in the foreground extinction. The random contributions are given by 
0.1 mag from the fit to the PNLF, 0.05 mag from the photometric zero point, 
and 0.05 mag from the filter calibration.
Combining all these errors quadratically, we estimate that the total error 
bar for the PNLF distance modulus is $\pm$0.2 mag. The PNLF distance 
modulus, 31.4, is smaller than the SBF distance modulus (31.9, according
to Tonry et al. 2001).

We would like to show that the use of the traditional analytical PNLF
with a universal cutoff at $M(5007)=-4.5$
(e.g., Ciardullo et al. 2002) gives a very similar distance estimate, as
demonstrated by Figure 15. The effect of sample size on the PNLF distance 
happens, if at all, only for a larger sample size, like that obtained for 
NGC 4697 by M\'endez et al. (2001). In that case the PNLF distance modulus
from simulated PNLFs was 30.1, which in our opinion is more reliable than
the smaller 29.9 obtained from the same data by Ciardullo et al. (2002) 
using the analytical PNLF. 

Once the sample size from Figure 14 is known, we can calculate the 
specific PN formation 
rate $\dot{\xi}$ in units of PNs yr$^{-1}$ $L_{\odot}$$^{-1}$, 

\begin{equation}
n_{\rm PN} = \dot{\xi} L_{\rm T} t_{\rm PN},
\end{equation}

where $n_{\rm PN}$ is the sample size, $L_{\rm T}$ 
is the total bolometric luminosity 
of the sampled population, expressed in $L_{\odot}$, and $t_{\rm PN}$ is the 
lifetime of a PN, for which 30,000 yr was adopted in the PNLF simulations. 
We have $B_{\rm T}$=11.67, $B-V$ = 0.93 (de Vaucouleurs et al. 1991), 
$A_{\rm B}$ = 0.47 (from data listed in 
NED, NASA/IPAC Extragalactic Database), and a 
bolometric correction of $-$0.8 mag (Buzzoni et al. 2006) from which we 
obtain an extinction-corrected apparent bolometric magnitude 9.47. Using a 
distance modulus of 31.4 and a solar $M_{\rm bol}$=4.72, we calculate the 
total luminosity of NGC 821 to be 4.6 $\times$ 10$^{10}$ $L_{\odot}$. 

Now we must consider that the statistically complete PN sample was built 
by excluding an elliptical region at the center of the galaxy.
We estimate that the excluded region of NGC 821 
contributes 60\% of the total luminosity, and we 
conclude that the luminosity effectively sampled is  
$L_{\rm T}$ = 1.8 $\times$ 10$^{10}$ $L_{\odot}$. Adopting 
$n_{\rm PN}$ = 2200 from Figure 12, we obtain 
$\dot{\xi}$ = (4 $\pm$ 2) $\times$ 10$^{-12}$ PNs 
yr$^{-1}$ $L_{\odot}$$^{-1}$. 

We can also express the PN formation rate as 
$\alpha = n_{\rm PN} / L_{\rm T}$.
Using that definition, we find log $\alpha$ = $-$6.91. An inspection of,
for example, Figure 12 in Buzzoni et al. (2006) shows that the $\alpha$ 
of NGC 821 is perfectly comparable, within the uncertainties, to that
of other similar elliptical galaxies.

\section{COMPARISON OF PNLF AND SBF DISTANCES}

Ciardullo et al. (2002) have reported a systematic discrepancy between 
distances determined from the PNLF and distances derived from the method 
of surface brightness fluctuations (SBF; Tonry et al. 2001). From a
comparison of distances to 28 galaxies, Ciardullo et al. reported that the 
PNLF distance moduli were smaller than the SBF moduli by 0.3 mag. This
situation has not been explained in any satisfactory way, in particular 
because both methods have been calibrated using Cepheid distances.

In the meantime, a few more PNLF distances for rather distant galaxies 
have been added: NGC 1344 in Fornax (Teodorescu et al. 2005), and the 
result we have just reported on NGC 821, which is the most distant 
galaxy for which a PNLF distance determination has been attempted. 
On the other hand, the SBF 
distance scale of Tonry et al. (2001) has been recalibrated 
in such a way that those ground-based SBF distances
have become smaller in modulus by 0.06 mag (Blakeslee,
private communication). For example, the SBF distance modulus of 
NGC 821 becomes 31.8. Furthermore, many SBF 
distances have been redetermined, and in all likelihood improved, 
using images obtained with the Hubble Space Telescope (HST) 
Advanced Camera for Surveys (ACS) (Blakeslee et al. 2009).
Therefore we have decided to make a new comparison of PNLF versus SBF 
distances, to put our results for NGC 821 and NGC 1344 in context.

We have used our own PNLF distances for NGC 4697, NGC 1344 and NGC 821. 
The PNLF distances for the other galaxies are taken from Feldmeier et al.
(2007) and Ciardullo et al. (2002). Concerning SBF distances, we have taken 
the Virgo and Fornax data, as well as NGC 4697, from Blakeslee et al. (2009).
All the others, including NGC 821, are from Tonry et al. (2001), corrected by
$-$0.06 mag. In Figure 16 we show, for a total of 23 galaxies, the difference
in distance modulus between the PNLF and SBF as a function of the SBF 
distance modulus. We might have expected the PNLF-SBF discrepancy to be 
slightly reduced, because of the SBF recalibration; but in fact it has become 
slightly larger, because the new ACS distances are a bit larger for several
galaxies with PNLF distance. For example, the SBF distance of 
NGC 4697 has gone from 30.29 to 30.49, and that of NGC 1344 from 31.42 
to 31.60. NGC 821 also shows a rather large discrepancy. So looking at 
SBF distance moduli between 31 and 32 mag, we get the impression that the 
PNLF-SBF discrepancy is now $-0.4$ mag.

Is there any evidence that the discrepancy increases with distance? 
We would agree with Ciardullo et al. (2002) that the evidence is perhaps
suggestive but not convincing, because there are a few galaxies at much 
shorter distances
(NGC 5128, M81) with as big a negative difference. More information will 
be needed at distance moduli between 28 and 30.
This possible dependence on distance deserves careful consideration, 
because a clear explanation for its existence could be contamination
by background galaxies like object 104 in Table 2, discussed before. If 
there were many of those, the corrected PNLF would become fainter, and a 
fit to the simulated PNLF would require a larger PNLF distance,
removing the discrepancy with the SBF distance. However, our kinematic
studies of NGC 1344 and NGC 821 
can be used to reject the idea. Such a numerous population of background 
galaxies, if interpreted as 5007 \AA \ emitters, would be expected to 
show a uniform distribution in velocity across the on-band filter 
transmission curve, which has a width in velocity of 3600 km s$^{-1}$.
In fact, this is exactly what happened in the spectroscopic investigation 
of a group of candidate Virgo intracluster PNs by Kudritzki et al. (2000), 
where those candidates were shown to be background Ly$\alpha$ emitters. 
In the cases of NGC 1344 and NGC 821, only one such contaminating 
source was found in each galaxy, which is not enough to affect our PNLF 
distance determinations, even if we assume that there are a few others 
at the ``right'' velocity to avoid detection. The only way to be 
completely sure is of course spectroscopic confirmation of the PN nature
of the brightest sources, by detecting a second emission line, which would 
require a large amount of big telescope time. There is a motivation, 
however: the desire to learn about the abundances of the PNs in these
distant elliptical galaxies. So perhaps the test will be made at some
future time. We should add that one such test has already been made on 
the not so distant galaxy NGC 4697 (11 Mpc, M\'endez et al. 2005). All 
the tested sources were confirmed as PNs.

Having rejected the background galaxy contamination theory, the best 
explanation for the
PNLF-SBF discrepancy is still extinction effects (Ciardullo et al. 2002).
Confirmation of this idea on the PNLF side will likewise require deep 
spectra of the brightest detected PNs. Again we can mention NGC 4697: 
the PN spectra of M\'endez et al. (2005) show low or moderate internal 
extinction,
of the order of 0.2 mag on average, which appears to be similar to the 
amount of internal extinction detected in PNs belonging to the PNLF 
calibrating galaxy M 31 by Ciardullo \& Jacoby (1999). At the present 
time we find no good reason to increase the PNLF distances, and the 
PNLF-SBF discrepancy remains unexplained.

\section{SUMMARY OF CONCLUSIONS}

We have been able to increase the number of PN detections and PN velocity
measurements in the elliptical galaxy NGC 821. A comparison with the 
detections reported by the PN.S group indicates that we have confirmed 
most of their detections. The velocities measured by the two groups, 
using different telescopes, spectrographs and slitless techniques, are 
in good agreement. This confirms the reliability of detections 
and the good quality of the radial velocities reported by both 
groups. We have built a combined sample of 167 PNs, and have 
confirmed the keplerian decline of the LOSVD reported previously.
The PN kinematics agree with absorption-line data within PN 
uncertainties. A dark matter halo may exist around this galaxy, but 
it is not needed to keep the PN velocities below the local escape 
velocity as calculated from the visible mass.

We have also confirmed the misalignment in rotation between outer 
PNs and inner stars reported by Coccato et al. (2009).

We have measured the $m$(5007) magnitudes of 145 PNs and produced 
a statistically complete sample of 40 PNs. The resulting PNLF was 
used to estimate the distance to NGC 821, which becomes the most 
distant galaxy with a PNLF distance determination. We have also 
estimated the PN formation rate. The PNLF distance 
modulus is smaller than the SBF distance modulus by 0.4 mag. Our 
kinematic information permits to rule out the idea that a shorter 
PNLF distance could be produced by the contamination of the PNLF 
by background galaxies with emission lines redshifted into the 
on-band filter transmission curve. The PNLF-SBF 
distance discrepancy remains unexplained.

%\section{Acknowledgments}
This work was supported by the National Science Foundation (USA) under 
grants 0307489 and 0807522. In our research we made use of the NASA/IPAC
Extragalactic Database (NED), which is operated by the Jet Propulsion 
Laboratory, California Institute of Technology, under contract with the 
National Aeronautics and Space Administration.
It is a pleasure to acknowledge the help 
provided by the Subaru staff, in particular the support astronomers 
Youichi Ohyama, Takashi Hattori and Kentaro Aoki. 
Our thanks to Lodovico Coccato, John Blakeslee, and the anonymous 
referee, for useful comments.

\clearpage

\figcaption[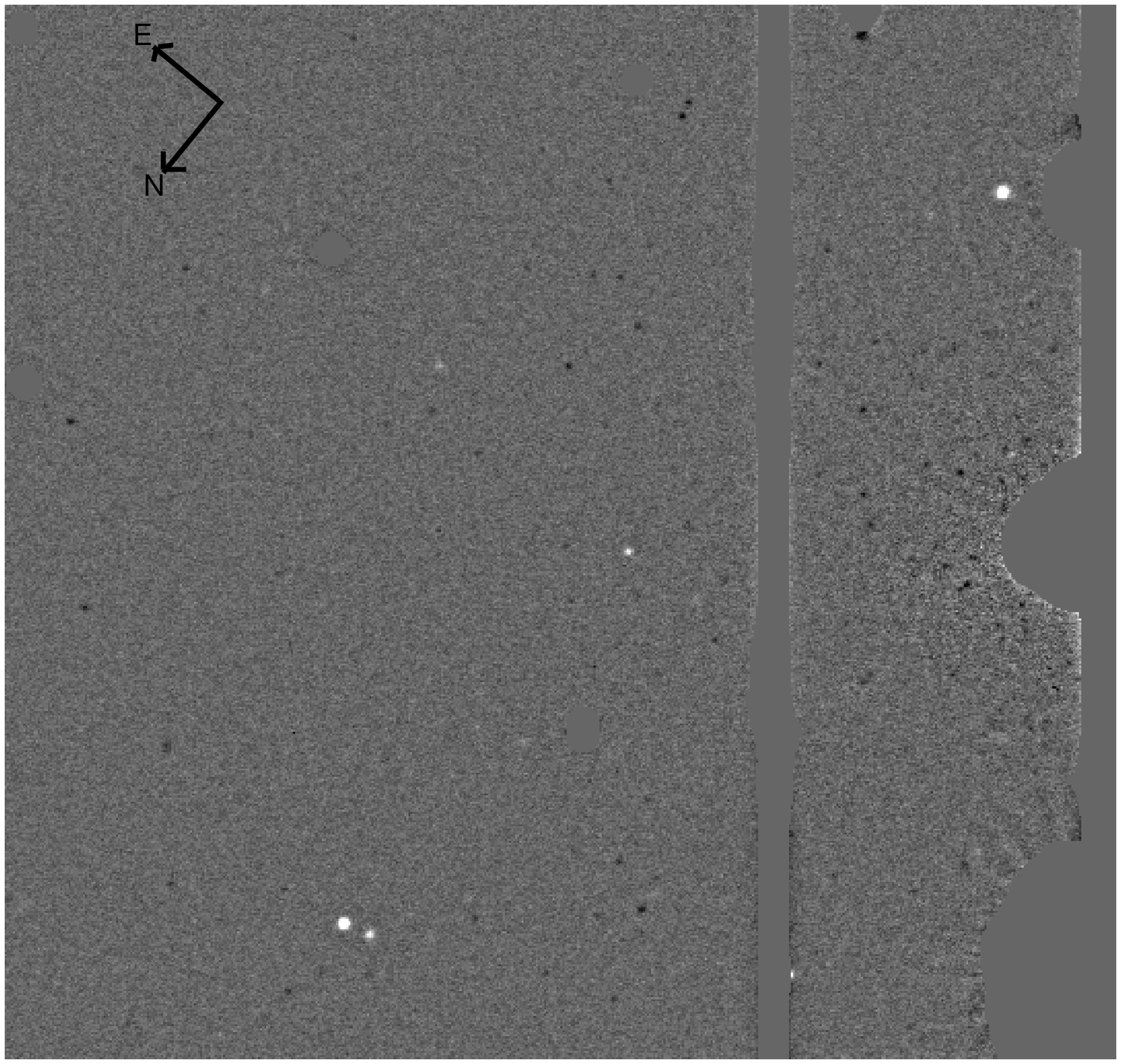]{NGC 821 difference image (on-off); part of Chip 2. 
The black dots are the PN candidates in the field. The sky area covered 
is 141 $\times$ 134 arcsec. The FOCAS images are specularly inverted 
relative to the sky.
\label{fig1}}

\figcaption[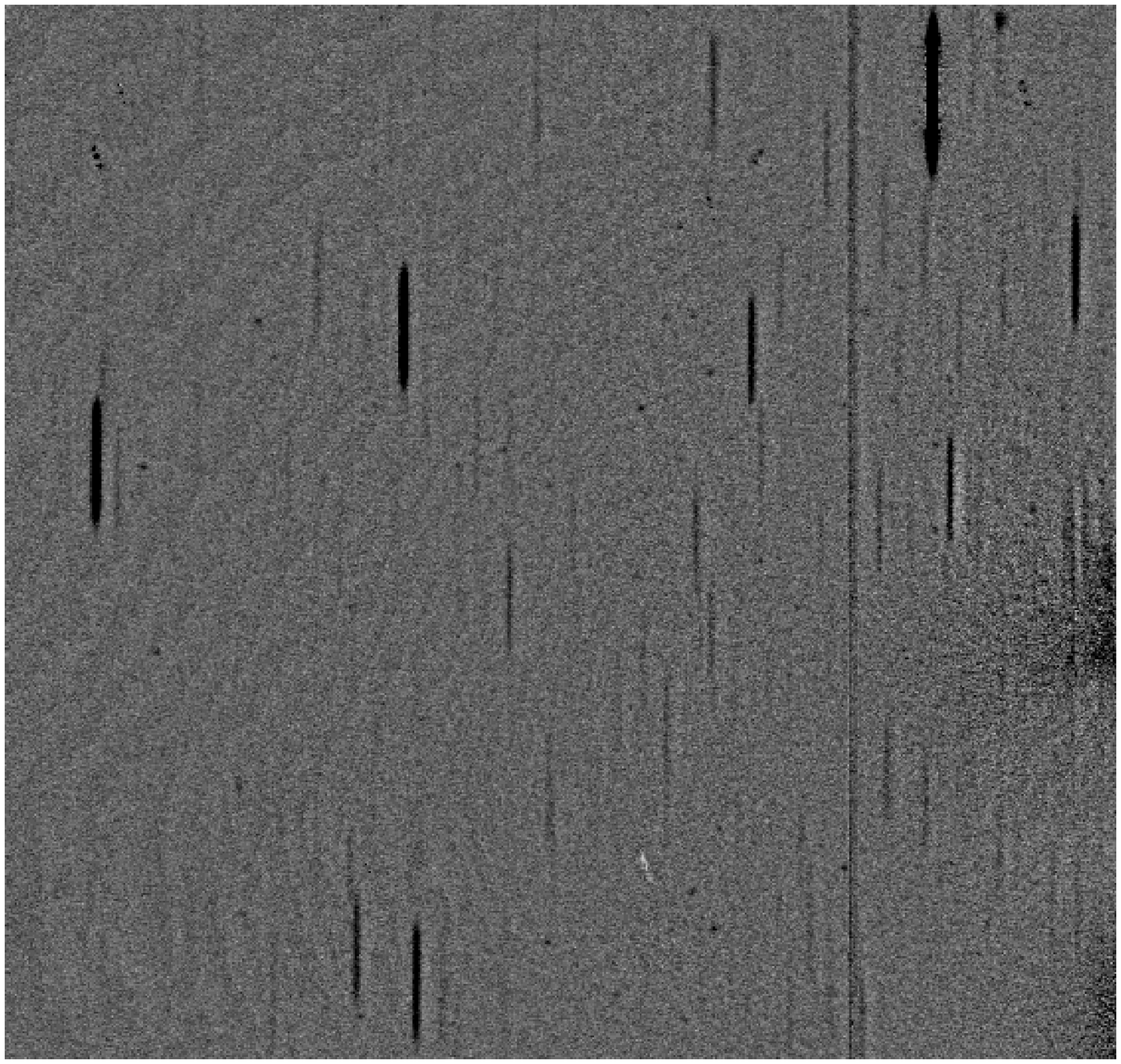]{NGC 821 difference image (unmedianed - medianed). 
This is the same part of Chip 2 shown in Figure 1. The spectra of 
continuum sources appear as vertical segments, while the spectra of
PNs remain as point sources.
\label{fig2}}

\figcaption[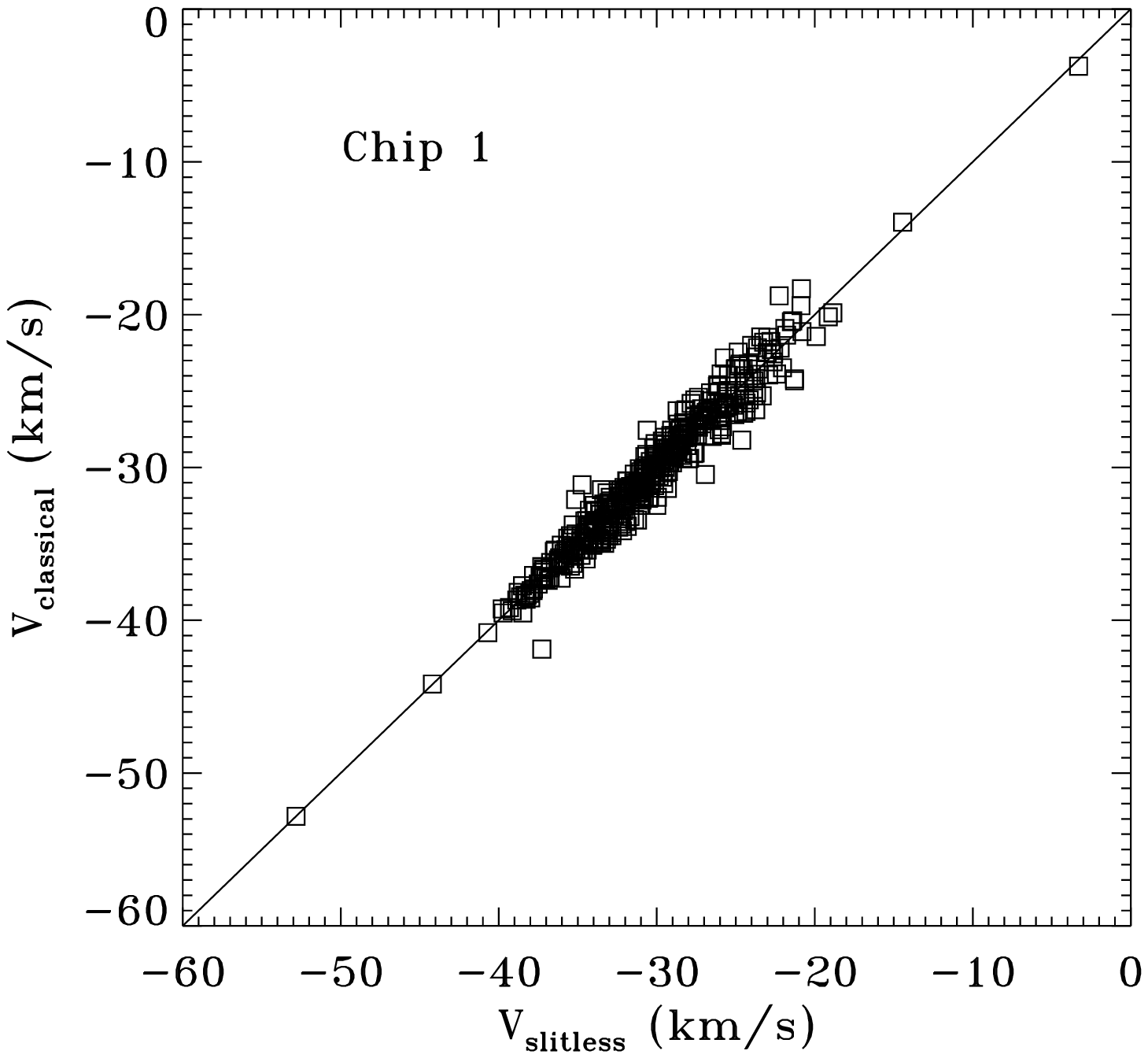]{Comparison of slitless vs slit (classical) radial 
velocities across NGC 7293 for both Chip 1 and Chip 2. This PN's 
heliocentric systemic 
radial velocity is $-$27 km s$^{-1}$, according to Meaburn et al. (2005).
\label{fig3}}

\figcaption[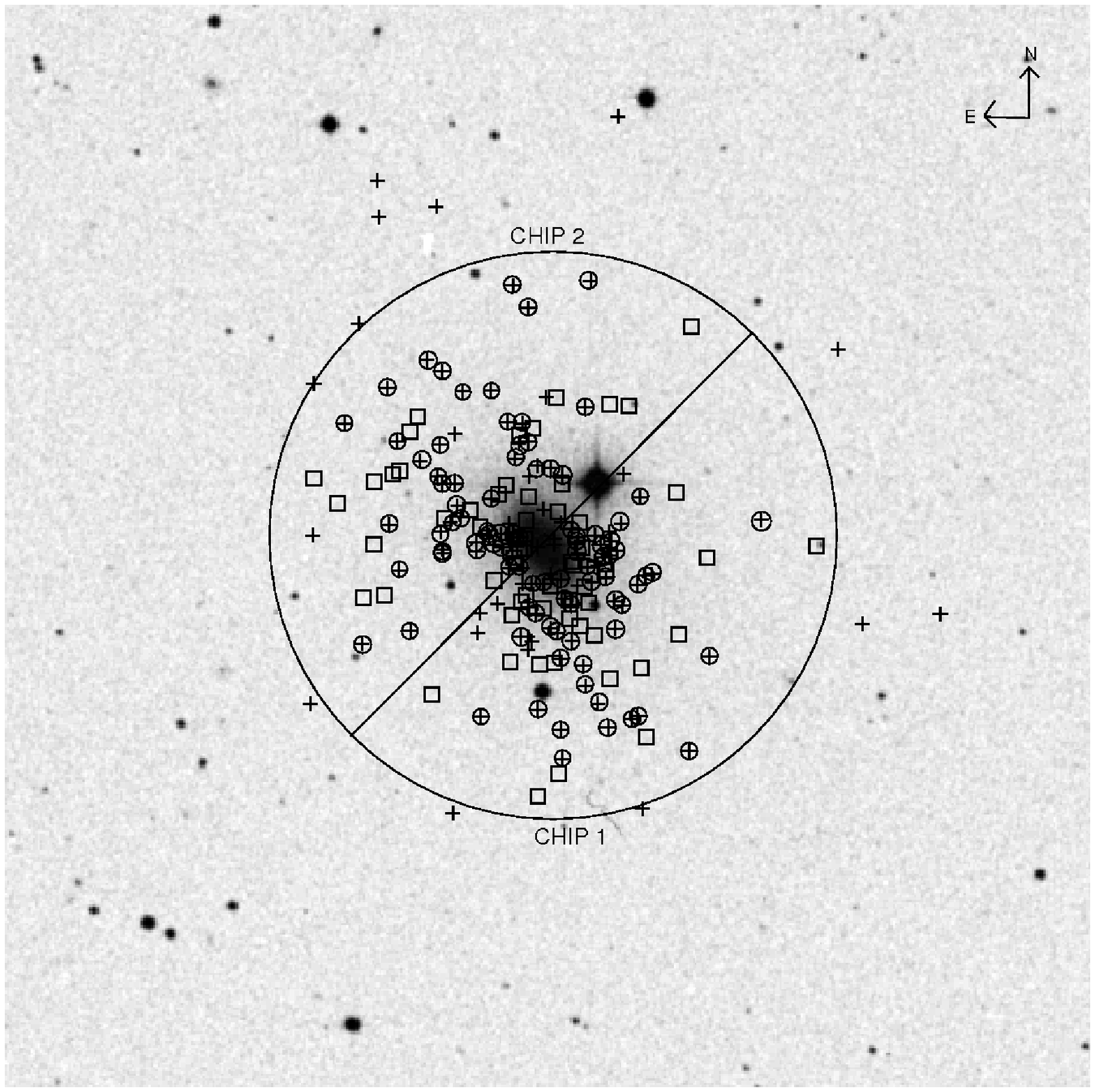]{Distribution of all PN candidates reported 
in NGC 821. The plus signs correspond to sources detected only with PN.S. 
The squares show sources detected only with Subaru + FOCAS. The circled 
plus signs show sources in common to both PN.S and FOCAS samples. We have
indicated the FOCAS field of view, divided into Chip 1 and Chip 2.
\label{fig4}}

\figcaption[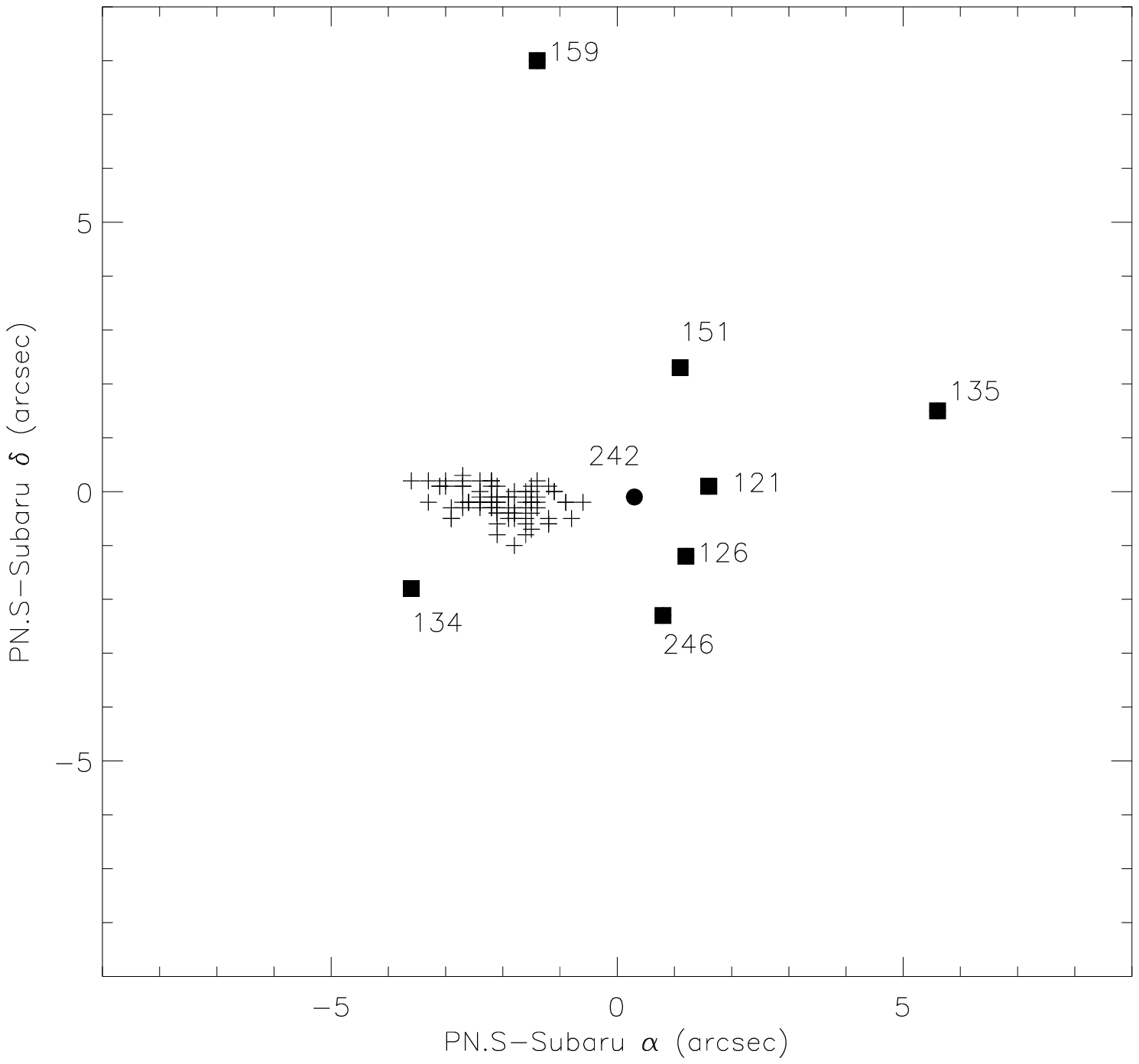]{Comparison of equatorial coordinates for 93 
PN candidates detected with both PN.S and FOCAS. We plot the difference 
in Declination as a function of the difference in Right Ascension.
We show 8 sources with too discrepant measurements. They are identified
with their FOCAS numbers from Table 2. We do not consider these 8 PN.S 
sources to be confirmed in our images. 
\label{fig5}}

\figcaption[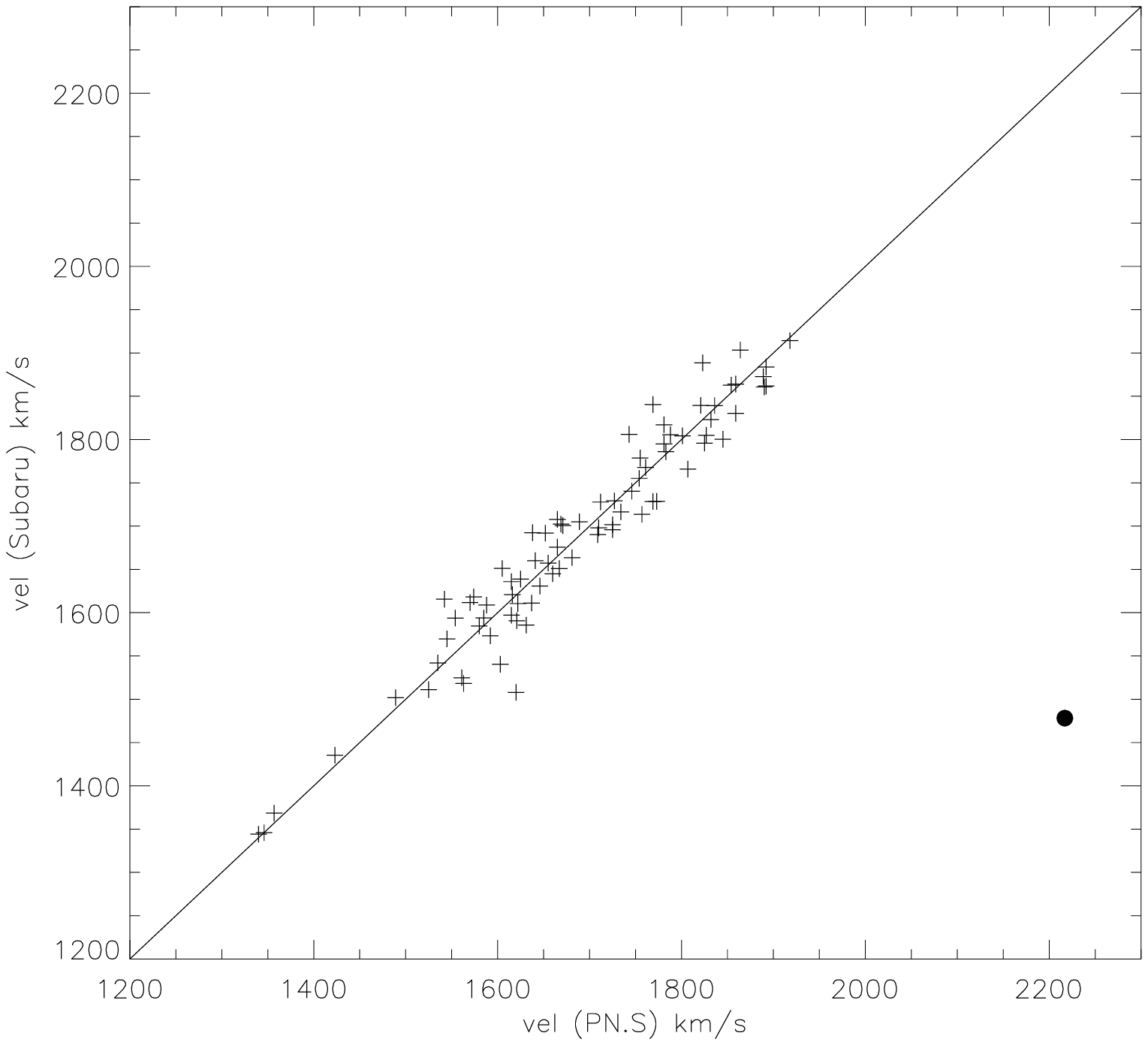]{Comparison of heliocentric velocities for 81 sources. 
We plot FOCAS velocities as a function of PN.S velocities.
The FOCAS object 242 is plotted as a filled circle. Eliminating object 
242 from the comparison, we obtain a standard deviation of 31 km s$^{-1}$. 
\label{fig6}}

\figcaption[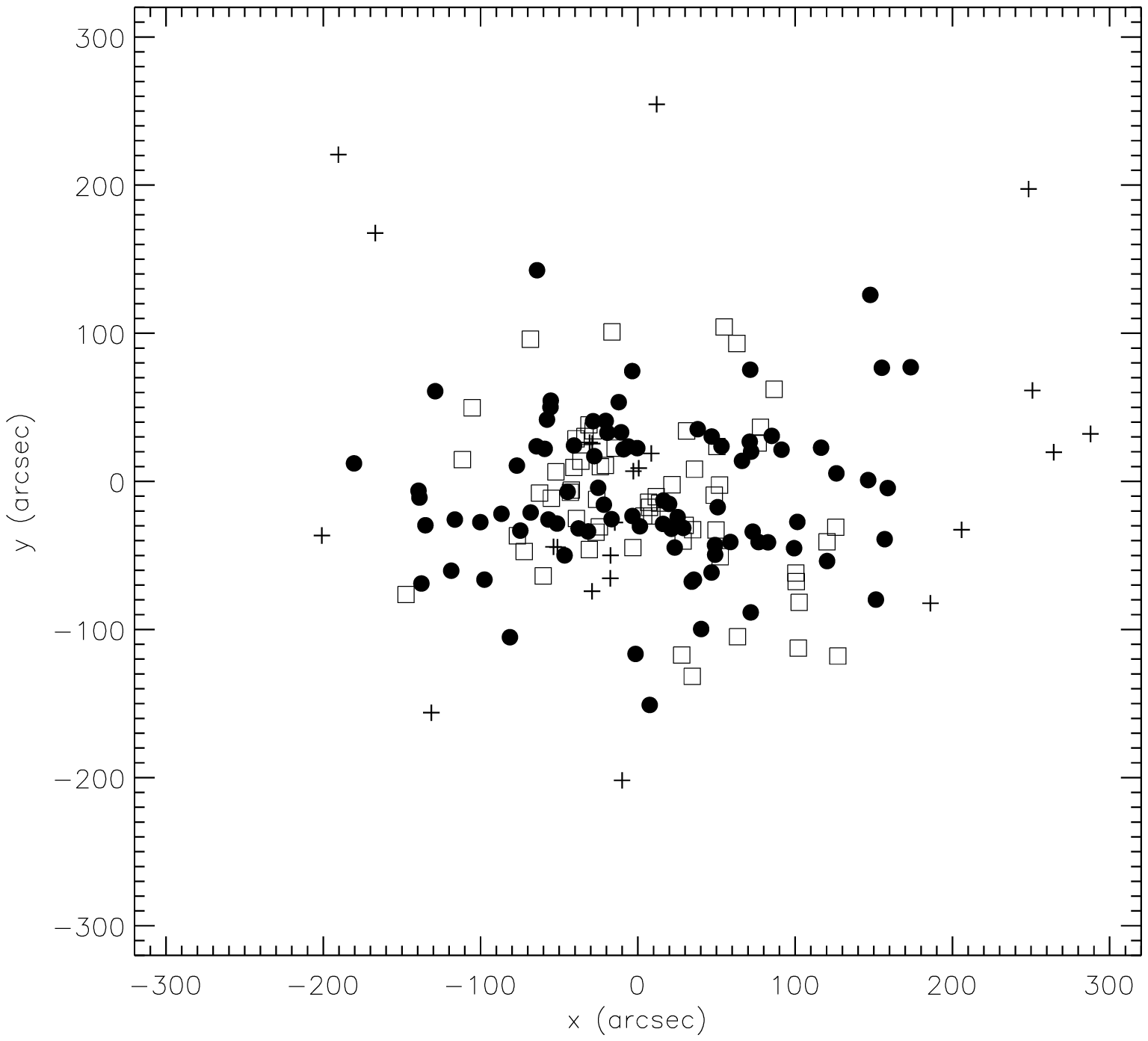]{Positions $x, y$ in arc seconds relative to the 
center of light of NGC 821, for the 168 PN candidates with 
measured velocities. The $x$-axis runs in the direction of increasing 
Right Ascension, along the major axis of NGC 821.
Plus signs represent PN.S detection only, open squares represent
FOCAS detection only, and filled circles are the PNs in common 
to both samples.
\label{fig7}}

\figcaption[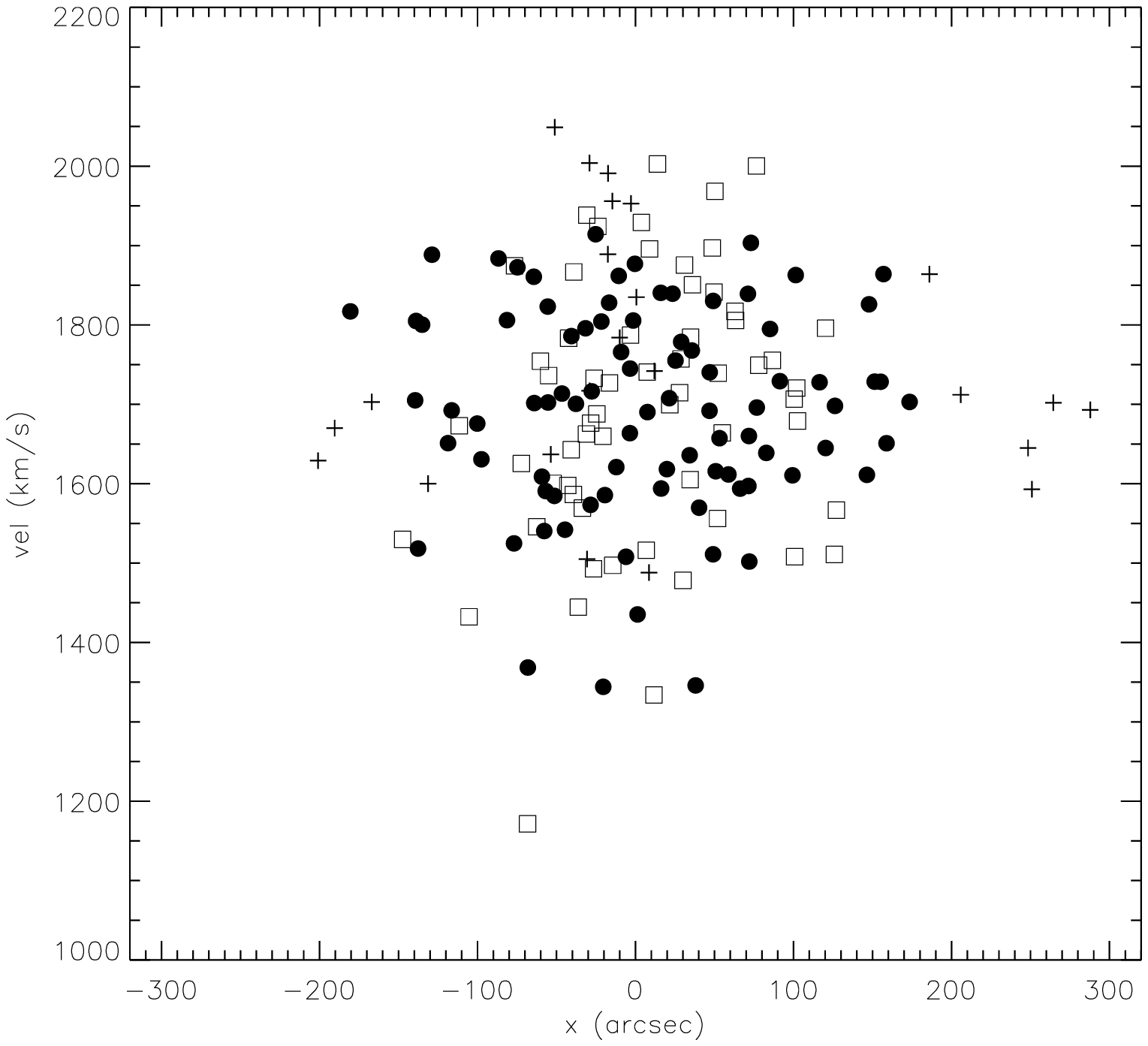]{Velocities of the 168 PN candidates as a function 
of their $x$-coordinates in arc seconds relative to the center of light 
of NGC 821. The symbols are the same as in figure 7. The isolated object
with the lowest velocity is FOCAS object 104 in Table 2.
\label{fig8}}

\figcaption[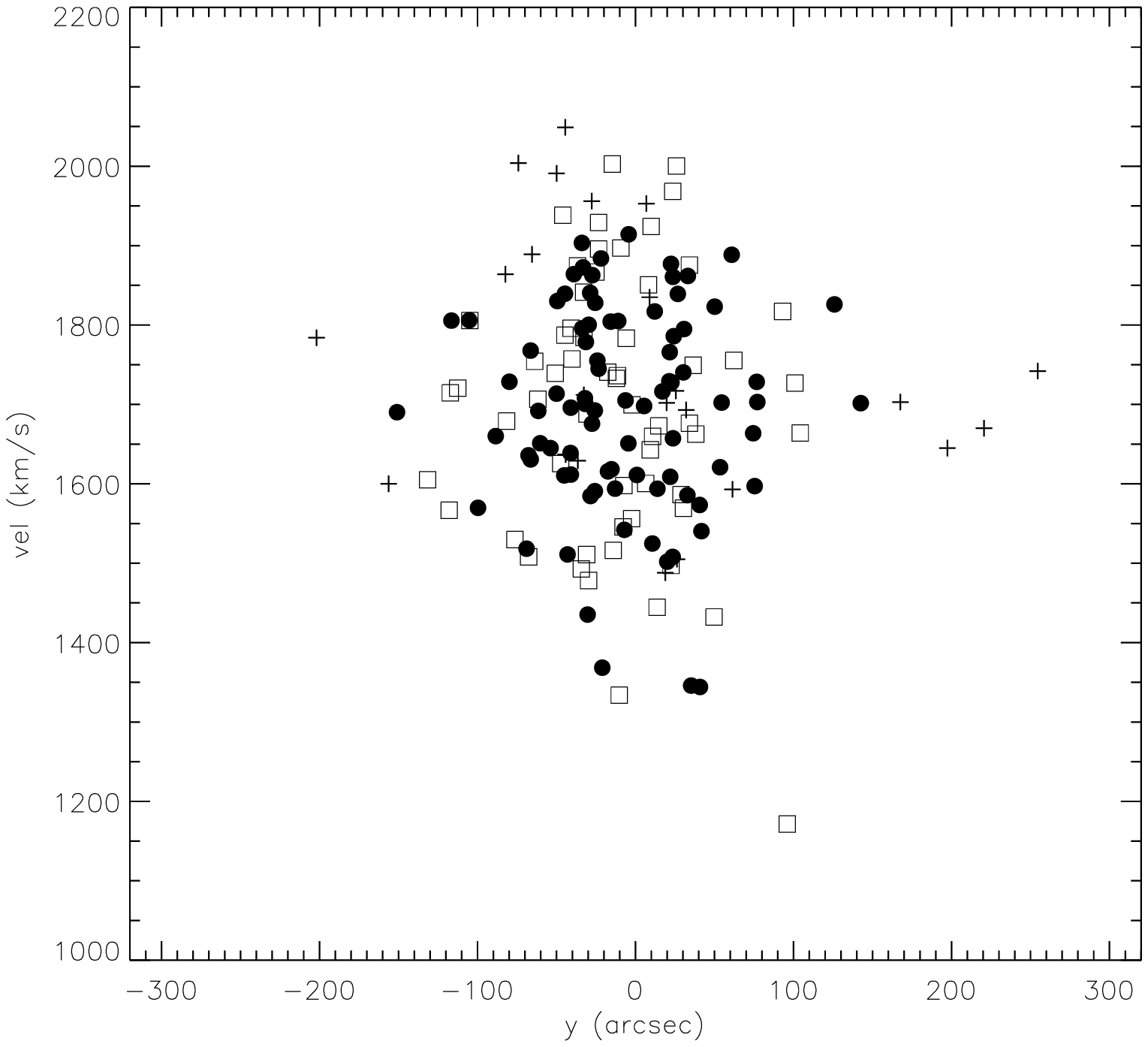]{Velocities of the 168 PN candidates as a function 
of their $y$-coordinates in arc seconds relative to the center of light 
of NGC 821. The symbols are the same as in figure 7. The isolated object
with the lowest velocity is FOCAS object 104 in Table 2.
\label{fig9}}

\figcaption[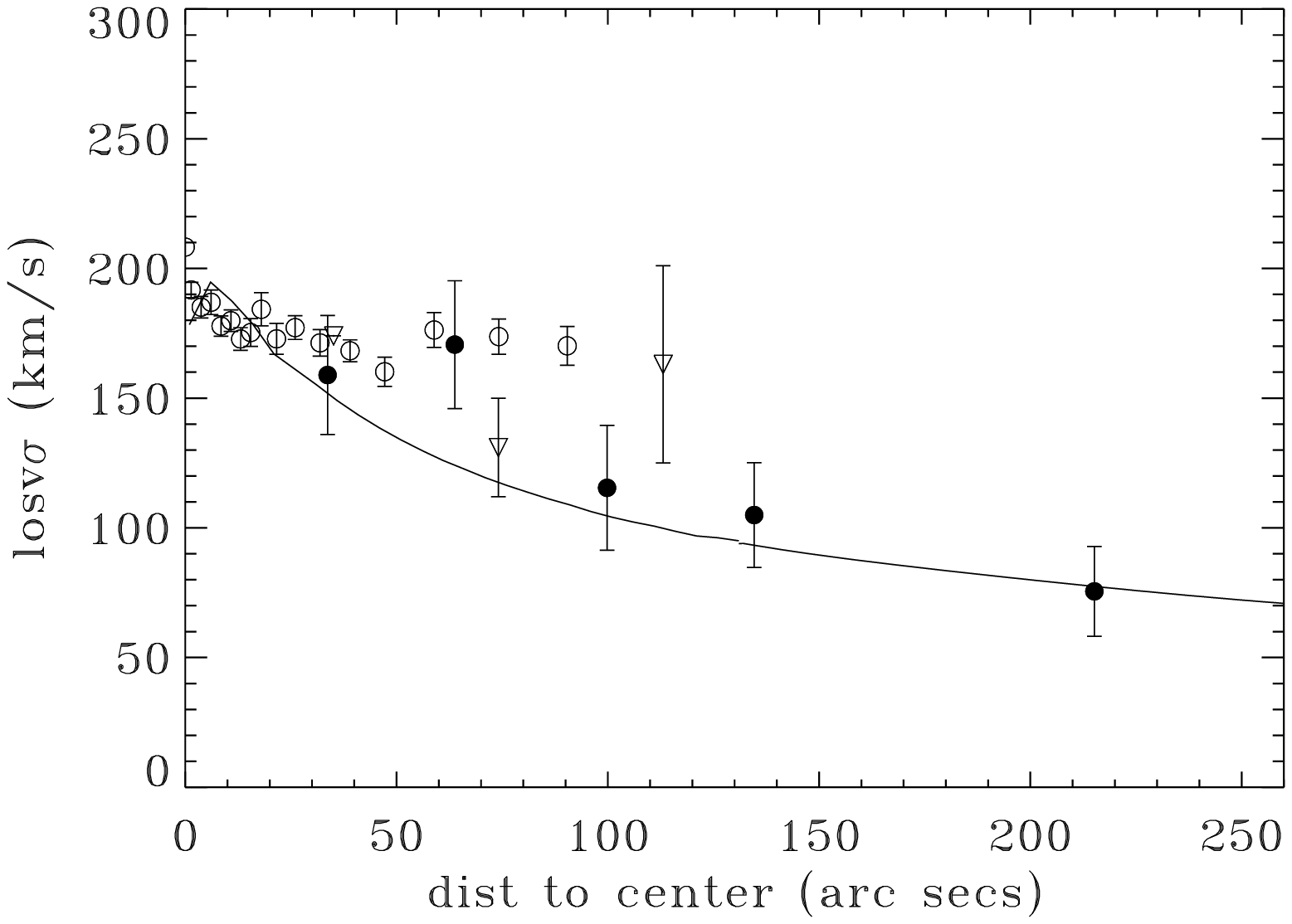]{Line-of-sight velocity dispersion as a function of 
projected angular distance to the center of NGC 821 for the FOCAS +
PN.S sample of 167 PNs (filled circles). 
The PNs were divided into 5 elliptical annuli as explained in the text.
Overplotted are Sauron data (triangles; Weijmans et al. 2009) and 
Forestell \& Gebhardt data (open circles). 
The solid line represents the analytical model of 
Hernquist (1990), with a constant $M/L$ ratio, a total mass of 
2 $\times$ 10$^{11}$ $M_{\odot}$, and $R_{\rm e}$= 39$''$.
\label{fig10}}

\figcaption[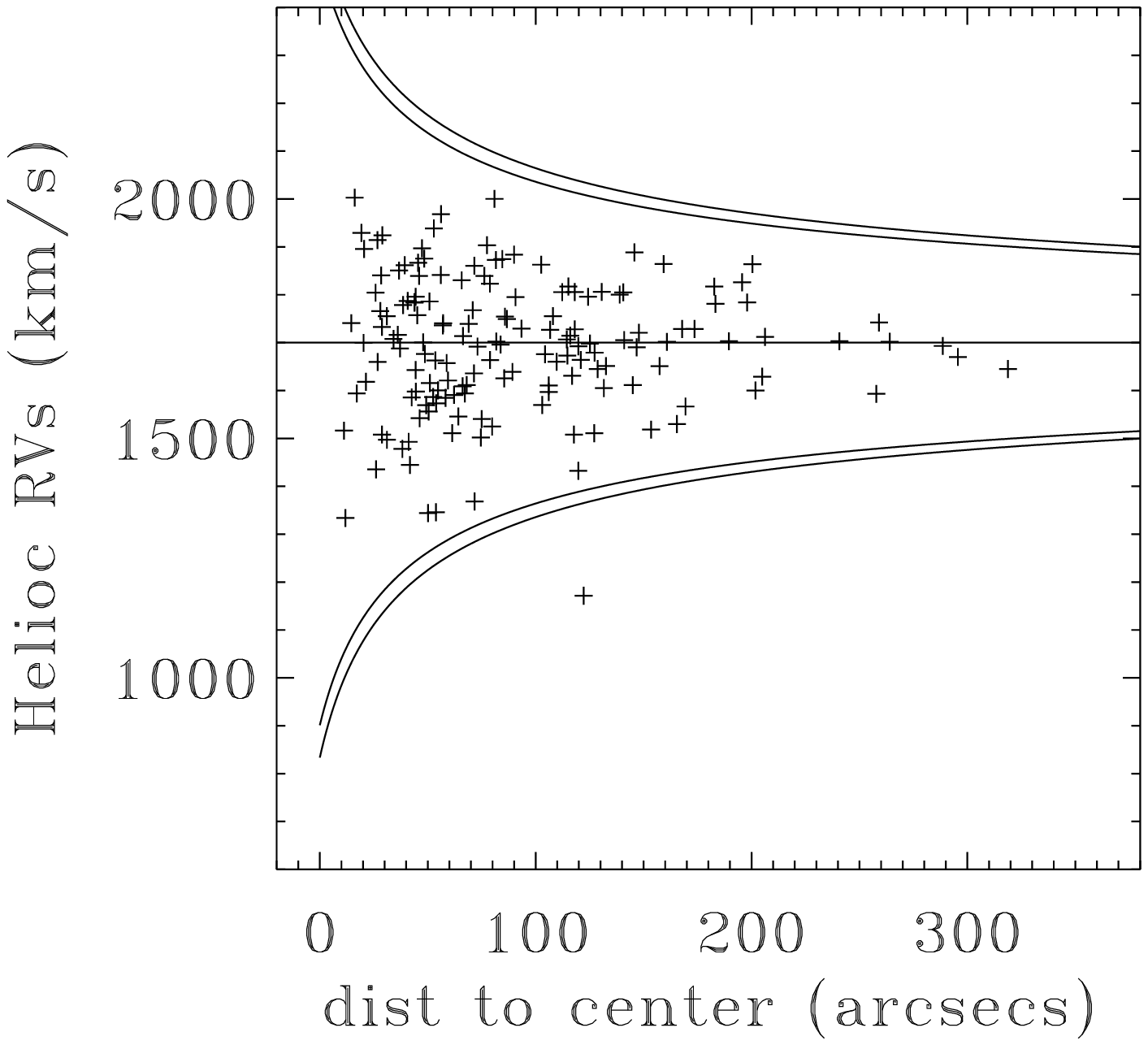]{Individual PN radial velocities plotted as a 
function of angular distance from the center of NGC 821. The solid 
lines are escape velocities for Hernquist models with total masses
2.0 x 10$^{11}$ $M_{\odot}$ (outer line) and 1.7 x 10$^{11}$ $M_{\odot}$ 
(inner line).
Note the outside position of the unrelated FOCAS object 104, which 
we have rejected as a PN. No other object has a velocity in excess
of the local escape velocity, even if we assume the lower mass. 
\label{fig11}}

\figcaption[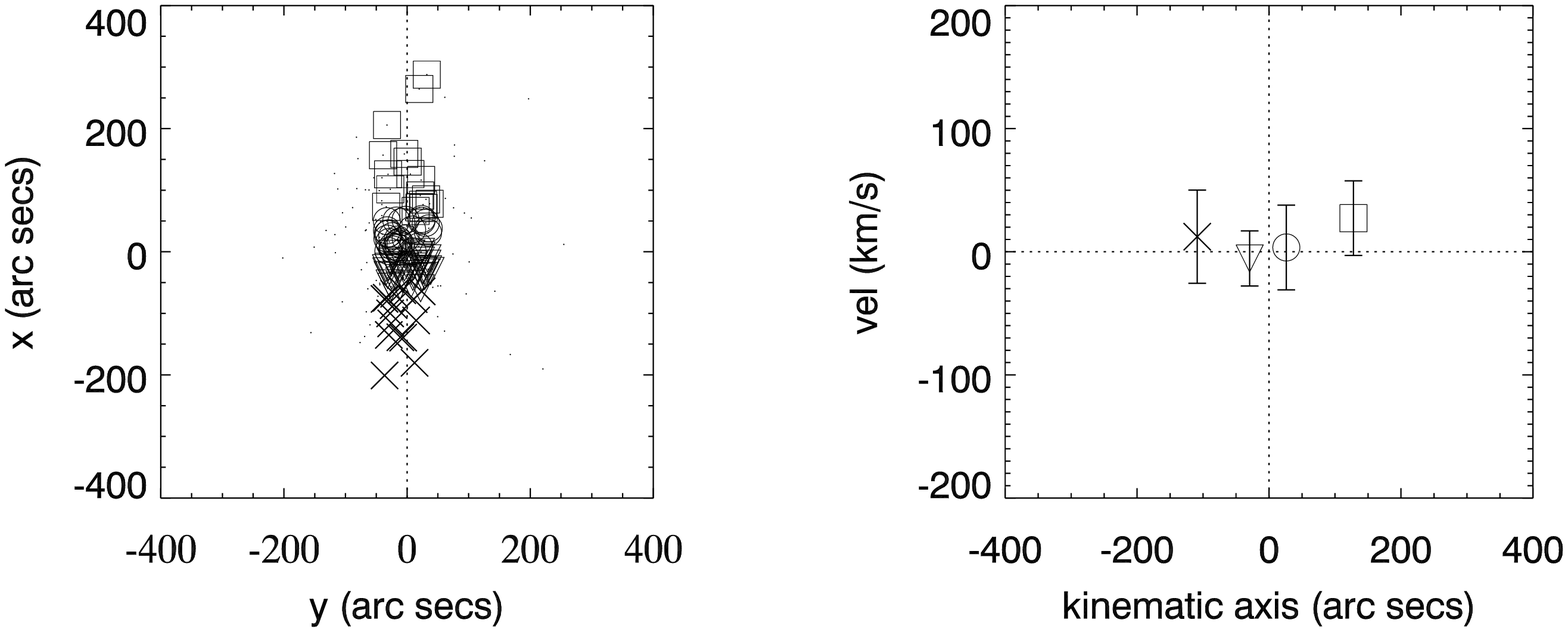]{Rotation of the PN.S + FOCAS sample of 167 PNs
along the photometric major axis. {\it Left}: distribution of PNs
in the sky, showing the same orientation as in Figure 3 of Coccato 
et al. (2009). The dotted line represents the major axis. We select all
PNs within $\pm 40''$ of the dotted line, and divide them in 4 bins.
Squares: $x > 60''$. Circles: $0'' < x < 60''$. 
Triangles: $-60'' < x < 0''$. Crosses: $x < -60''$.
{\it Right}: Average velocity for each of the four bins defined in the 
left figure. There is no significant rotation in this direction.
\label{fig12}}

\figcaption[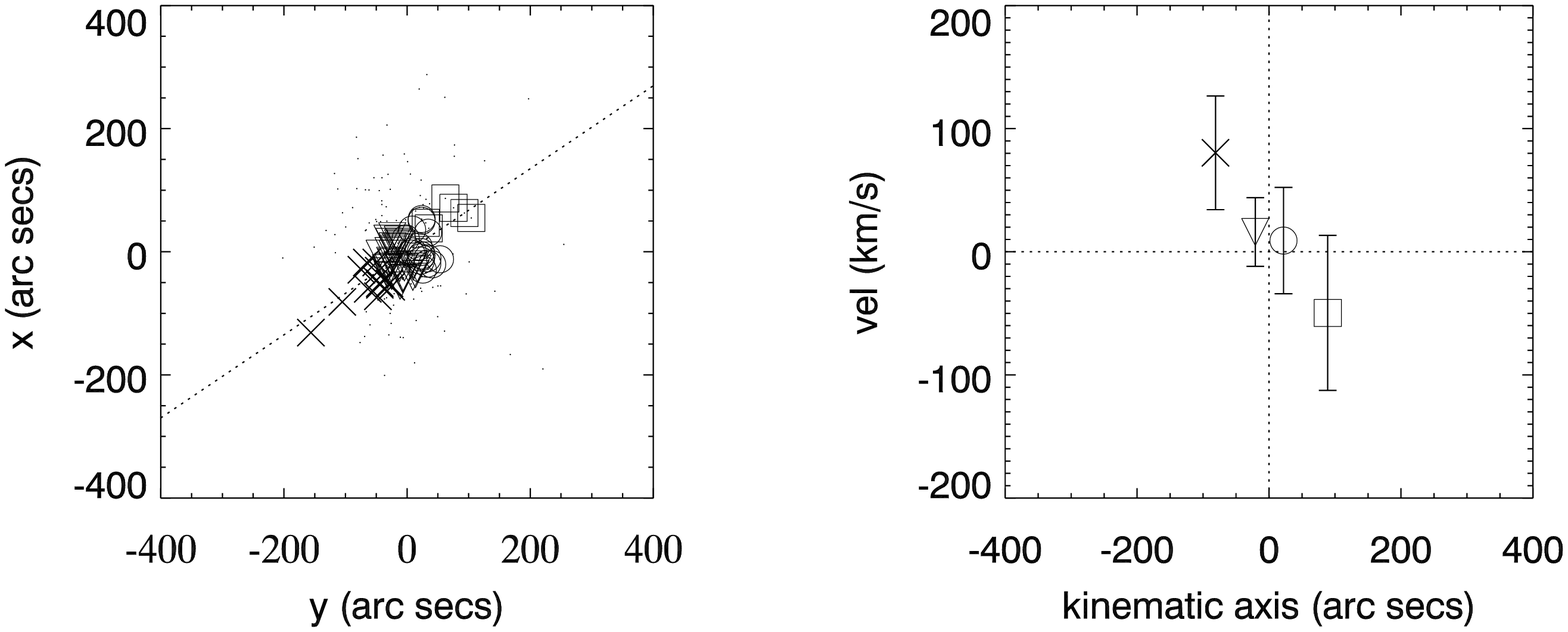]{Rotation of the PN.S + FOCAS sample of 167 PNs
along an axis inclined 56 degrees with respect to the photometric 
major axis. {\it Left}: distribution of PNs
in the sky, showing the same orientation as in Figure 3 of Coccato 
et al. (2009). The dotted line represents the inclined axis. We select all
PNs within $\pm 40''$ of the dotted line, calculate their $X$ coordinates 
with respect to the new axis, and divide them in 4 bins.
Squares: $X > 50''$. Circles: $0'' < X < 50''$. 
Triangles: $-50'' < X < 0''$. Crosses: $X < -50''$.
{\it Right}: Average velocity for each of the four bins defined in the 
left figure. The sense of rotation agrees with that reported by 
Coccato et al. (2009).
\label{fig13}}

\figcaption[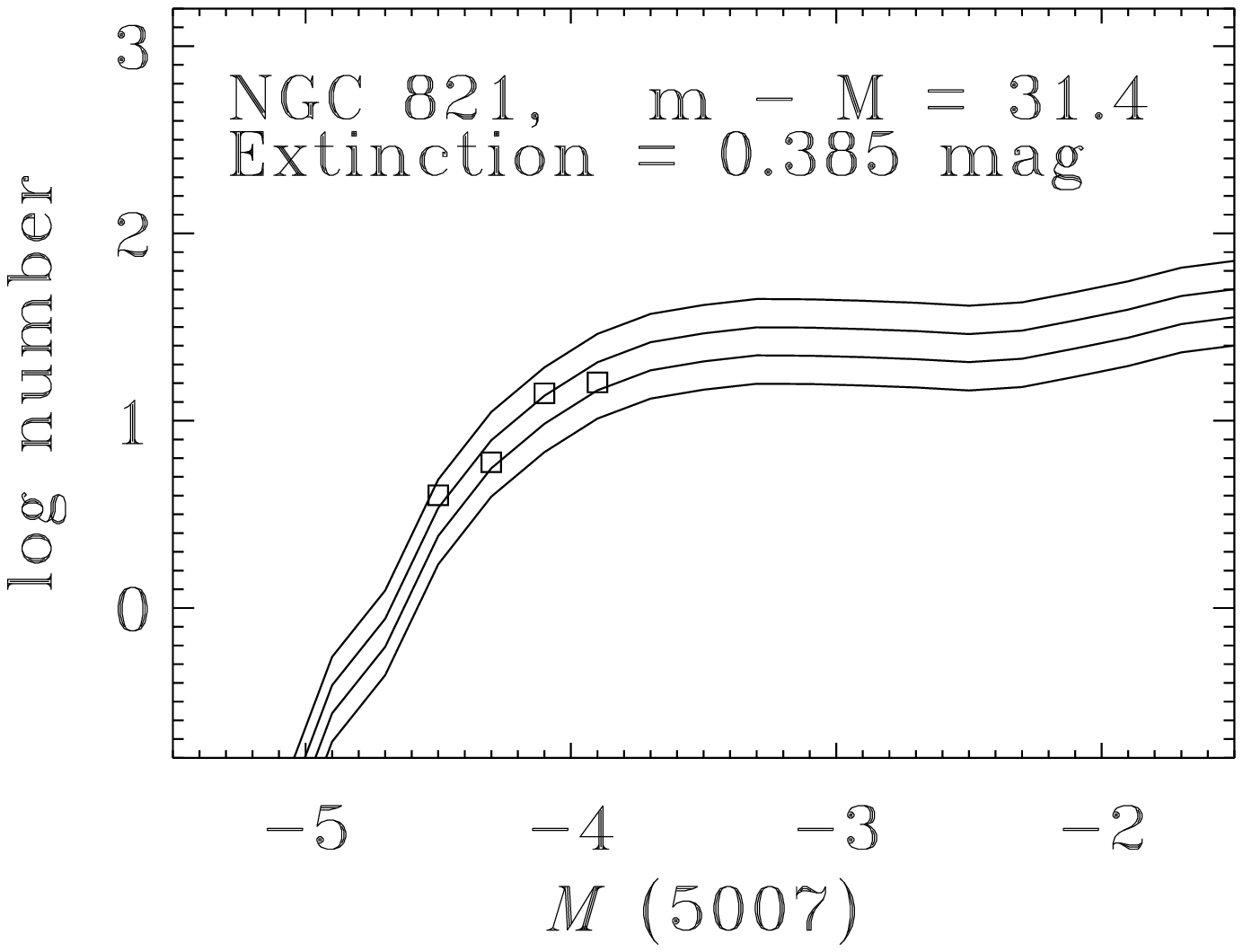]{Observed [O III] $\lambda$5007 PNLF of NGC 821 
(squares), with the statistically complete sample of 40 PNs binned 
into 0.2 mag intervals. The apparent magnitudes m(5007) have been 
transformed into absolute magnitudes M(5007) by adopting an extinction 
correction of 0.385 mag and a distance modulus m - M = 31.4. 
The 4 lines are PNLF simulations (M\'endez and Soffner 1997) for 
4 different total PN population sizes: 1200, 1700, 2400, and 3400 PNs. 
We estimate the best-fit sample size to be 2200. From this sample size 
it is possible to estimate the PN formation rate (see text).
\label{fig14}}

\figcaption[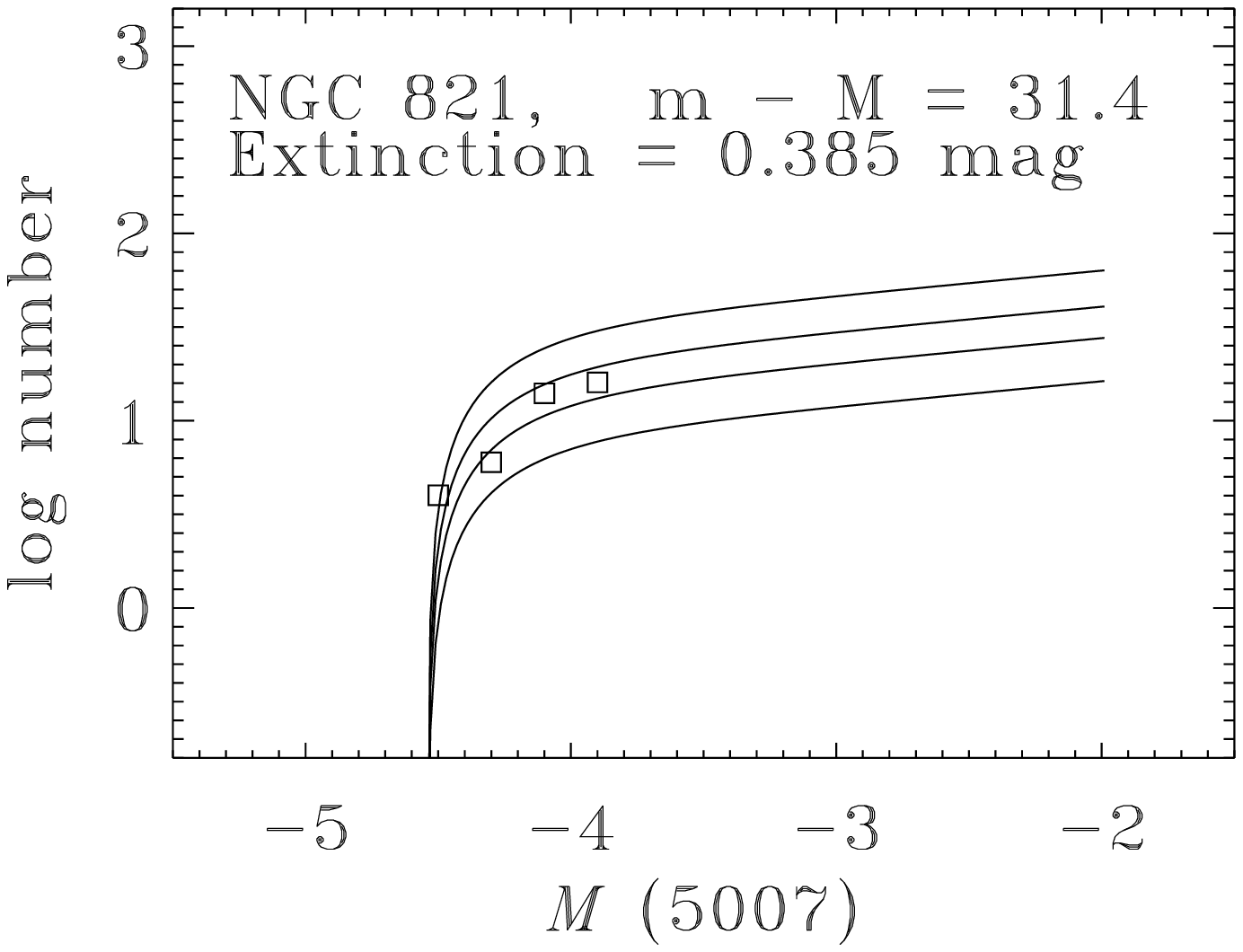]{Observed PNLF now compared with the analytical 
representation of the PNLF, using a universal cutoff at $-$4.5. 
The best-fit distance modulus is again 31.4.
\label{fig15}}

\figcaption[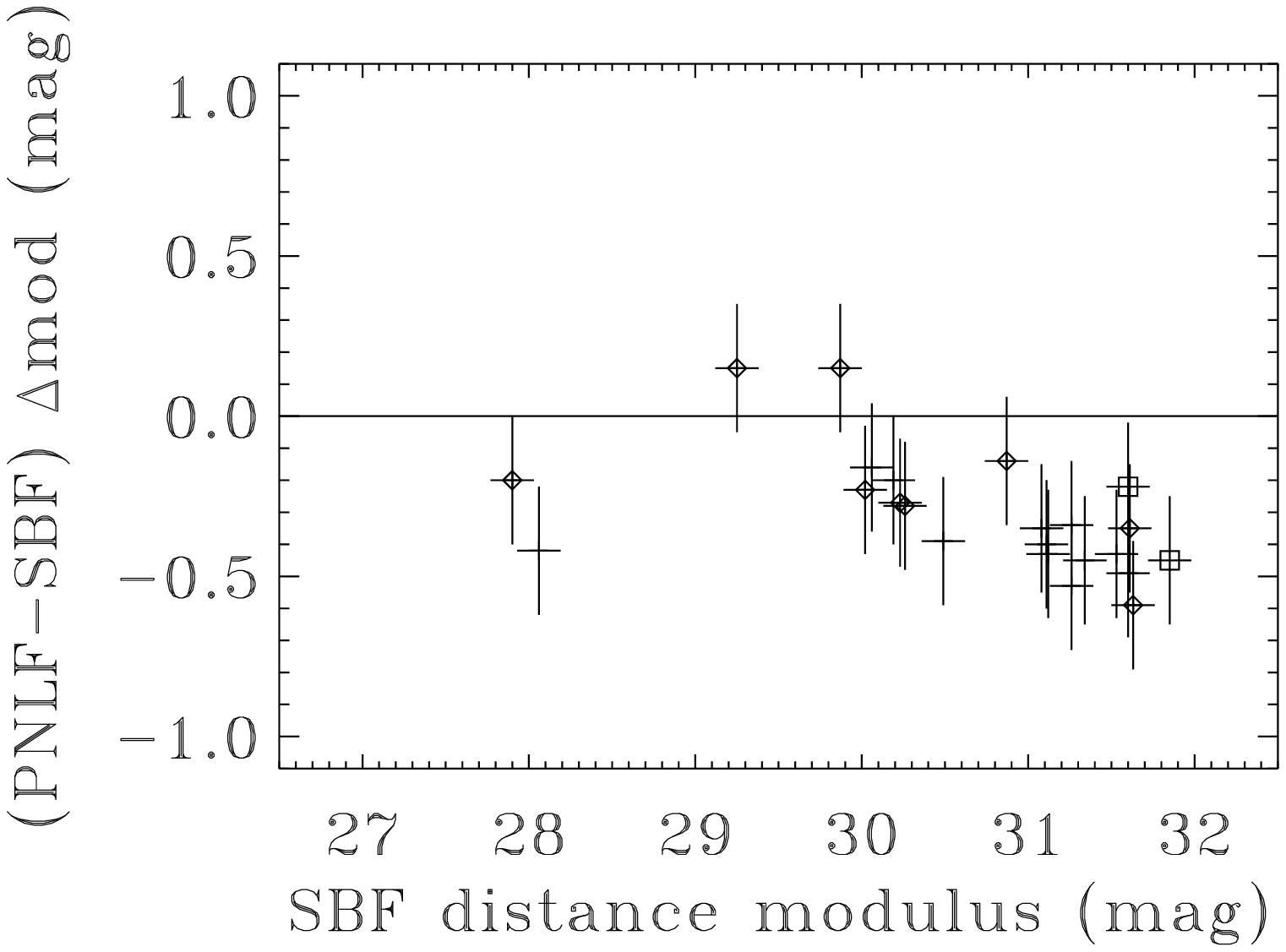]{Difference between PNLF and SBF distance modulus
plotted as a function of SBF distance modulus, for 23 galaxies. 
Non-elliptical galaxies are plotted as diamonds, while for ellipticals 
we show just the error bars. The two galaxies near $m-M$=28 are NGC 5128 
and M 81. The only two galaxies with a positive $\Delta$mod are NGC 4258
and NGC 3115. The agglomeration of galaxies between $m-M$ 31 and 32 
is dominated by the Virgo and Fornax clusters. Our recent PNLF additions, 
ellipticals NGC 1344 in Fornax and NGC 821, are indicated as squares.
\label{fig16}}

%%%%%%%%%%%%%%%%%%

\clearpage

\begin{figure}
\figurenum{1}
\epsscale{1.0}
\plotone{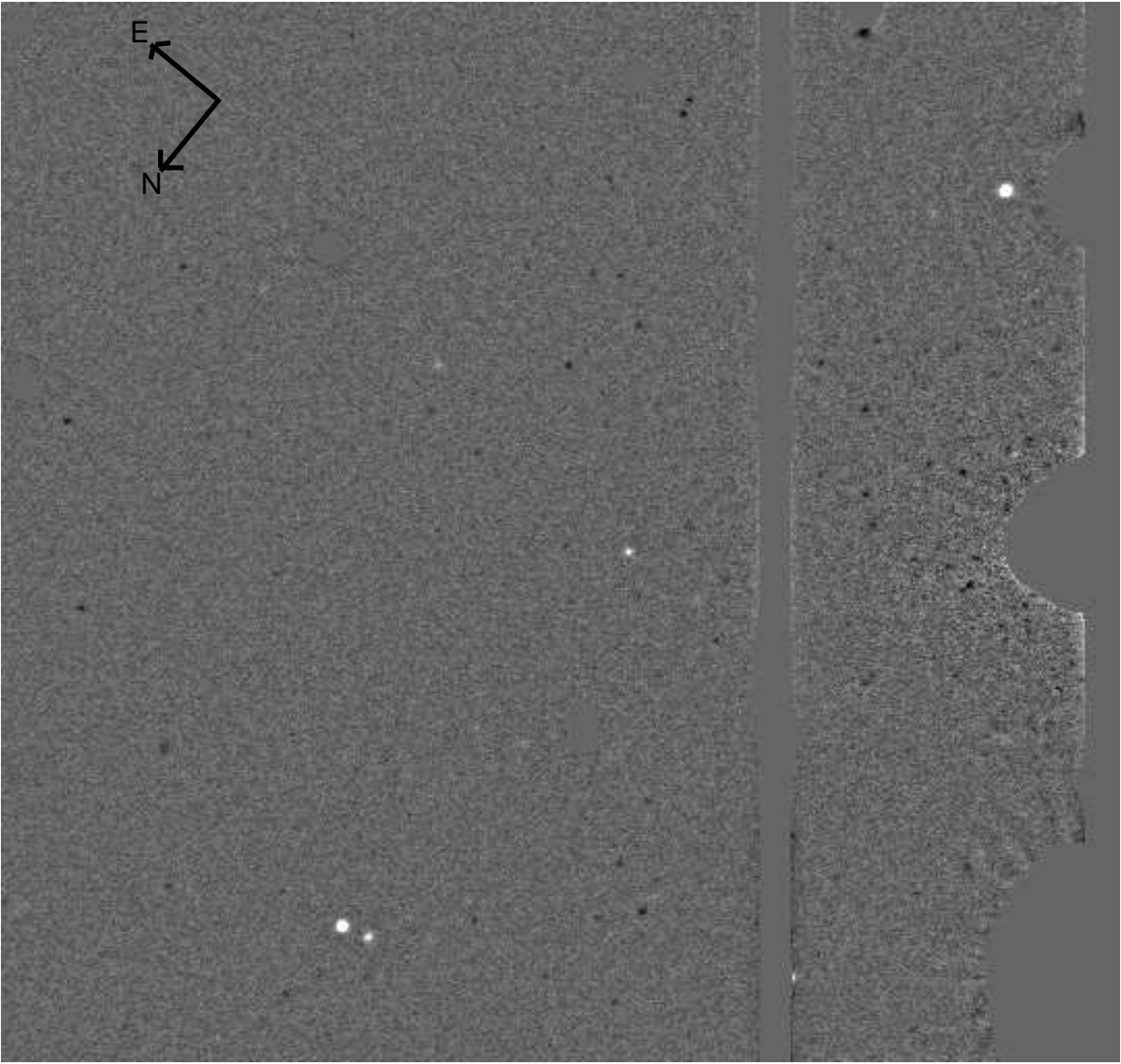}
\caption{NGC 821 difference image (on-off); part of Chip 2. 
The black dots are the PN candidates in the field. The sky area covered 
is 141 $\times$ 134 arcsec. The FOCAS images are specularly inverted 
relative to the sky.
}
\end{figure}

\begin{figure}
\figurenum{2}
\epsscale{1.0}
\plotone{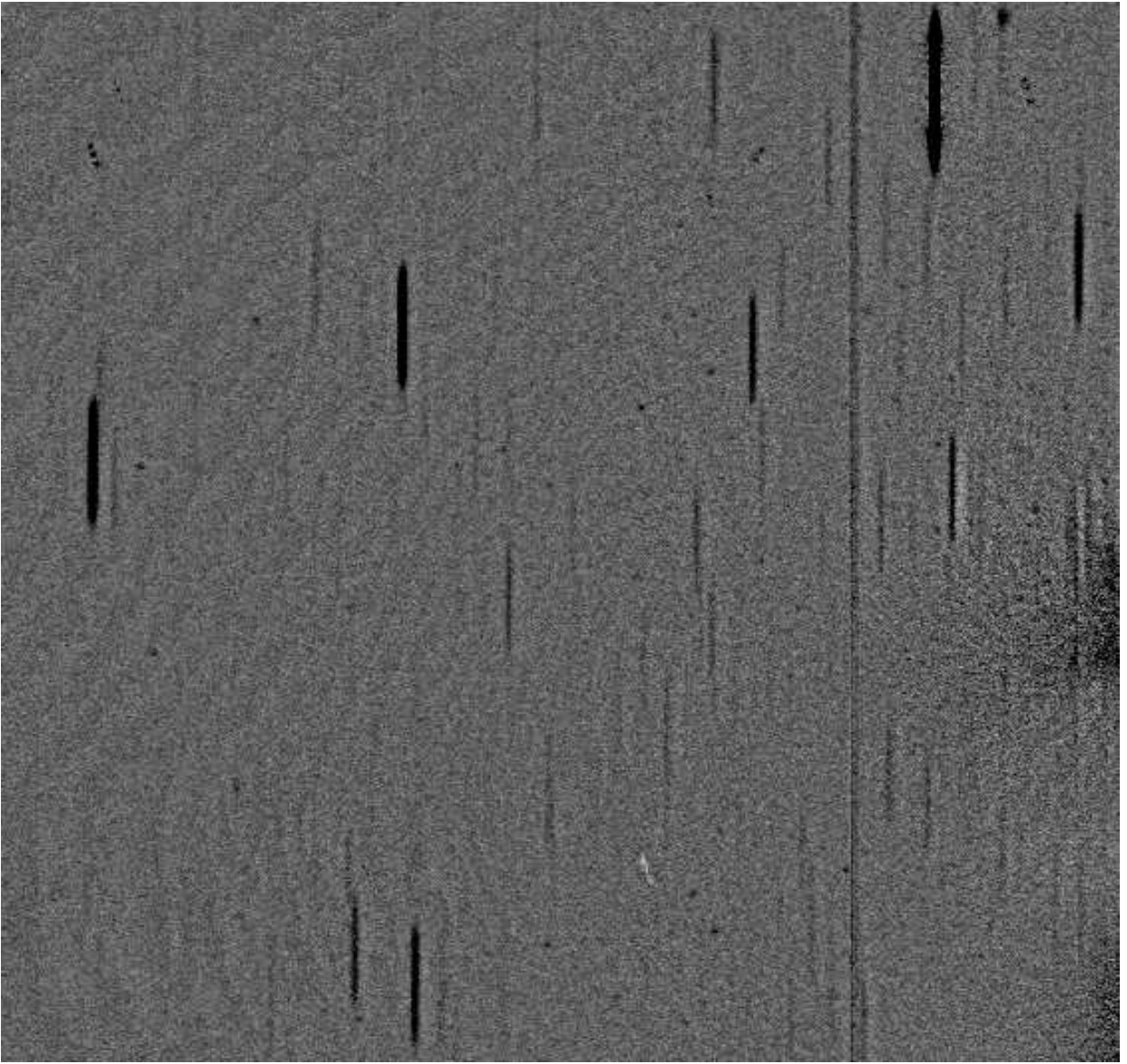}
\caption{NGC 821 difference image (unmedianed - medianed). This 
is the same part of Chip 2 shown in Figure 1. The spectra of 
continuum sources appear as 
vertical segments, while the spectra of PNs remain as point sources.
}
\end{figure}

\begin{figure}
\figurenum{3}
\epsscale{1.24}
\plottwo{f03a.ps} {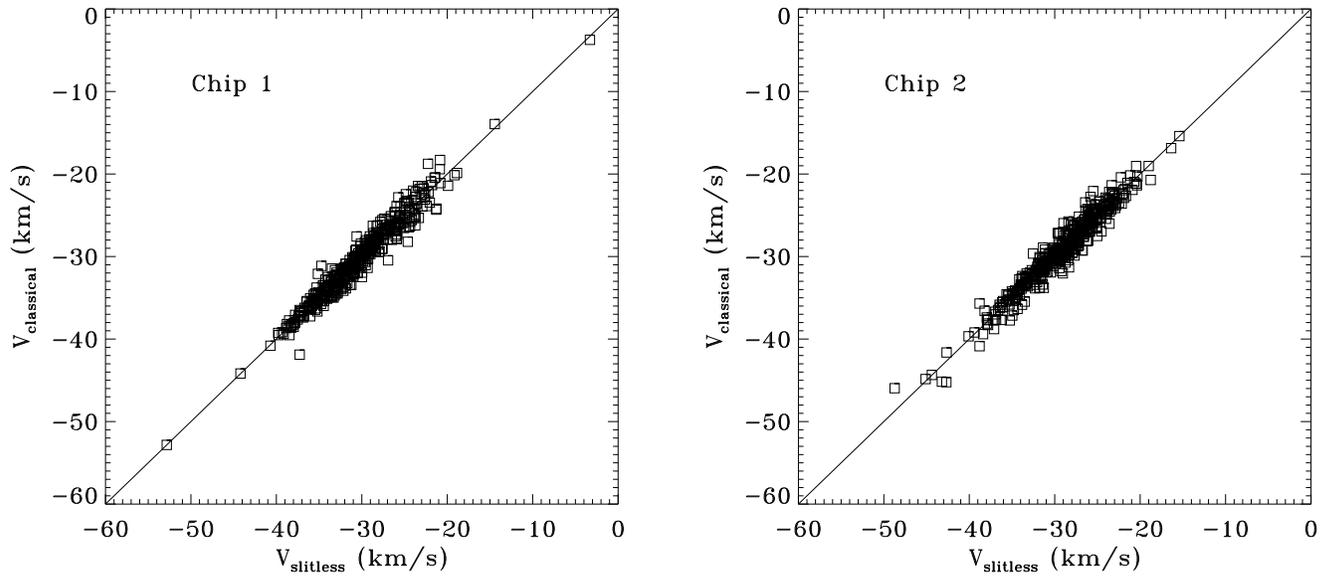}
\caption{Comparison of slitless vs slit (classical) radial 
velocities across NGC 7293 for both Chip 1 and Chip 2. This PN's 
heliocentric systemic 
radial velocity is $-$27 km s$^{-1}$, according to Meaburn et al. (2005).
}
\end{figure}

\begin{figure}
\figurenum{4}
\epsscale{1.0}
\plotone{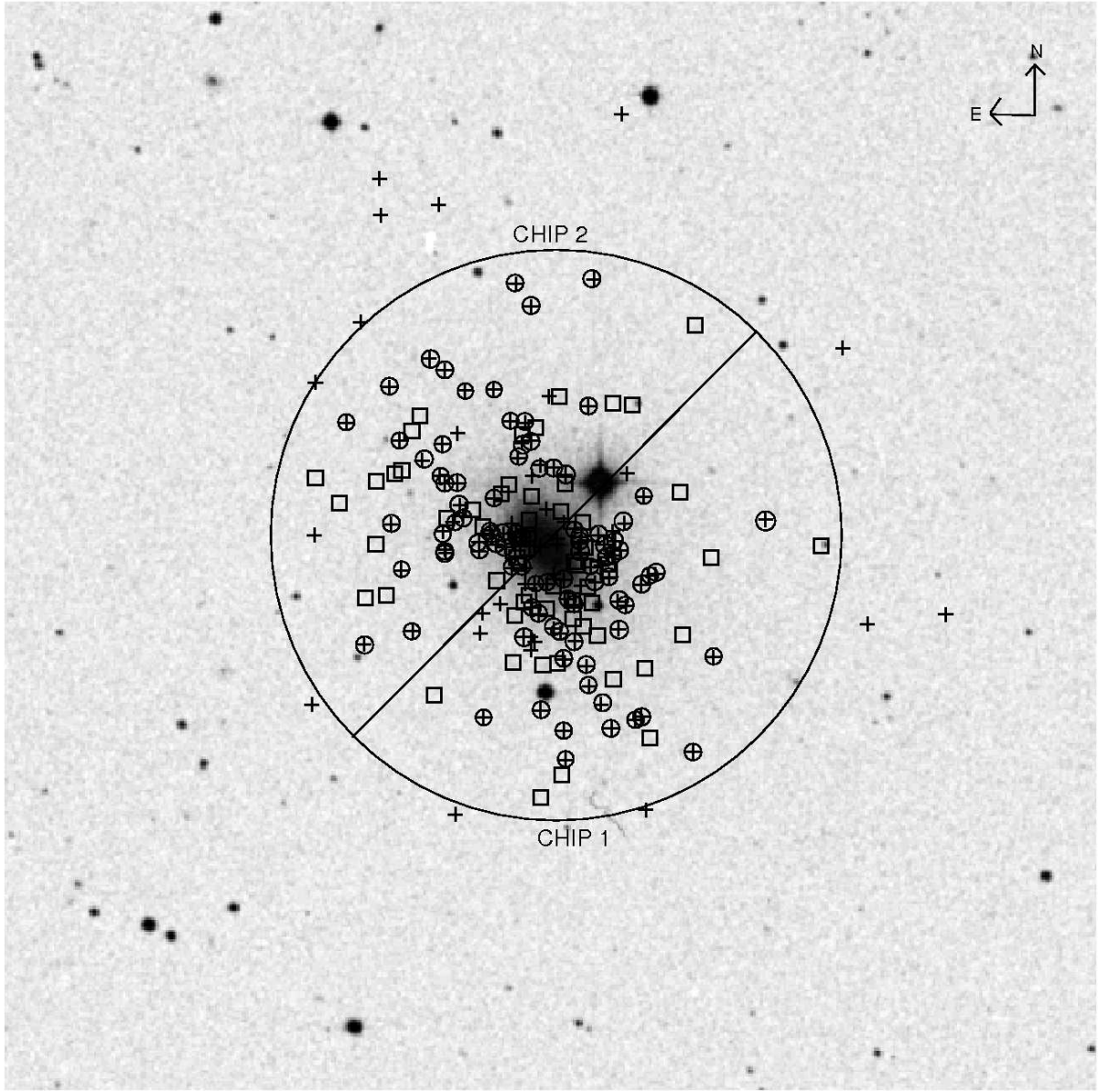}
\caption{Distribution of all PN candidates reported 
in NGC 821. The plus signs correspond to sources detected only with PN.S. 
The squares show sources detected only with Subaru + FOCAS. The circled 
plus signs show sources in common to both PN.S and FOCAS samples. We have
indicated the FOCAS field of view, divided into Chip 1 and Chip 2.
}
\end{figure}

\begin{figure}
\figurenum{5}
\epsscale{1.0}
\plotone{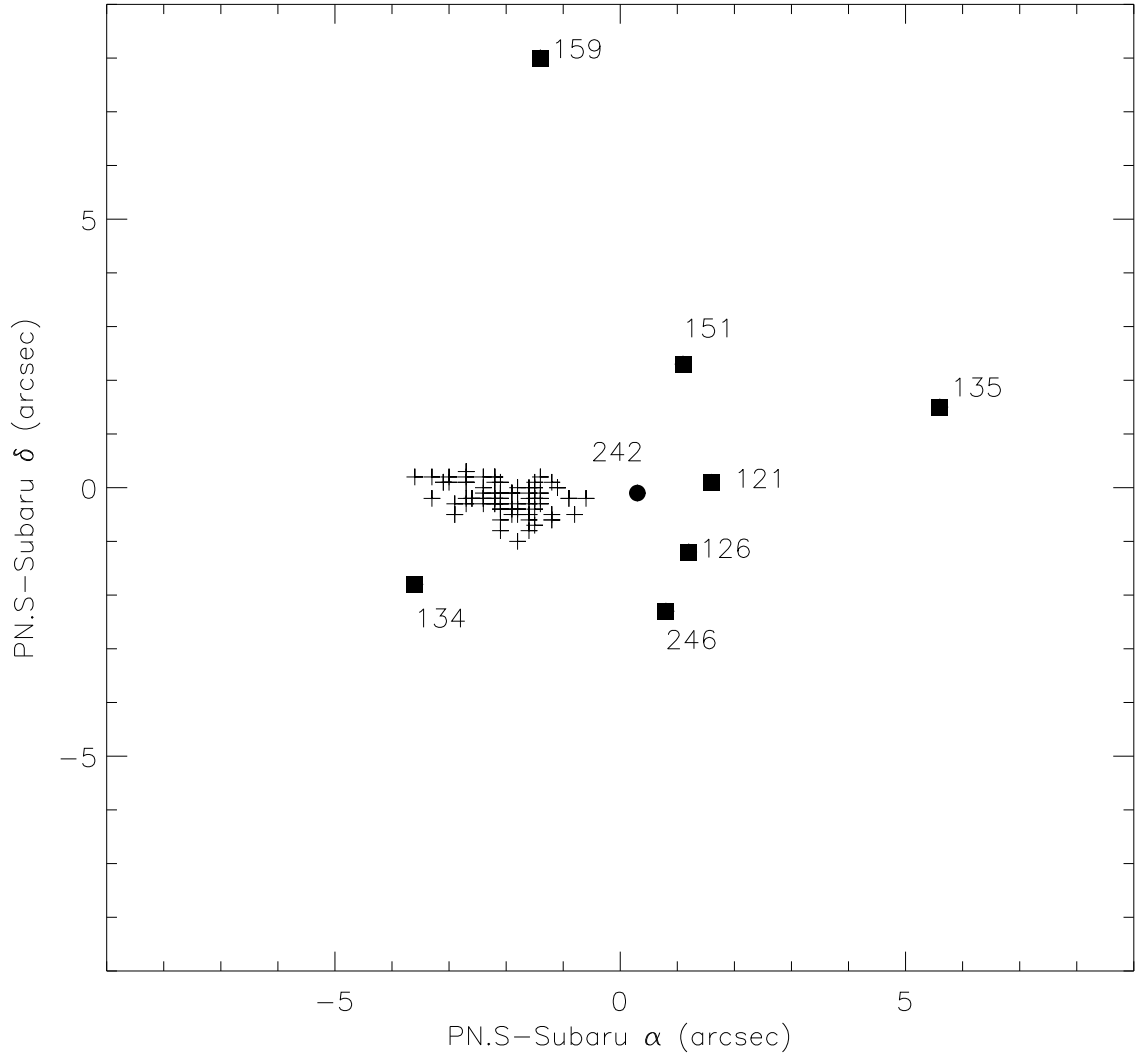}
\caption{Comparison of equatorial coordinates for 93 
PN candidates detected with both PN.S and FOCAS. We plot the difference 
in Declination as a function of the difference in Right Ascension.
We show 8 sources with too discrepant measurements. They are identified
with their FOCAS numbers from Table 2. We do not consider these 8 PN.S 
sources to be confirmed in our images.
}
\end{figure}

\begin{figure}
\figurenum{6}
\epsscale{1.0}
\plotone{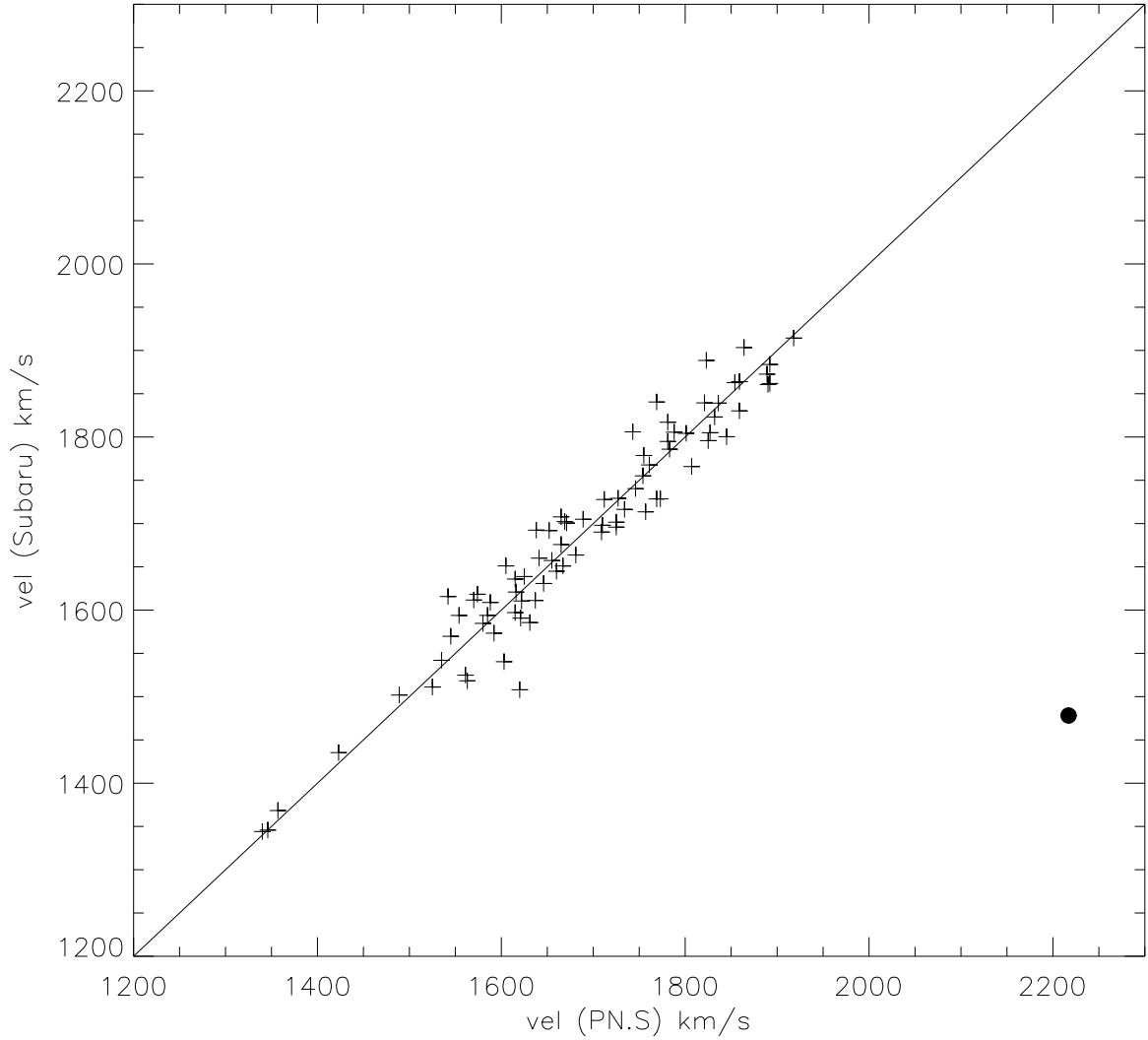}
\caption{Comparison of heliocentric velocities for 81 sources. 
We plot FOCAS velocities as a function of PN.S velocities.
The FOCAS object 242 is plotted as a filled circle. Eliminating object 
242 from the comparison, we obtain a standard deviation of 31 km s$^{-1}$.
}
\end{figure}

\begin{figure}
\figurenum{7}
\epsscale{1.0}
\plotone{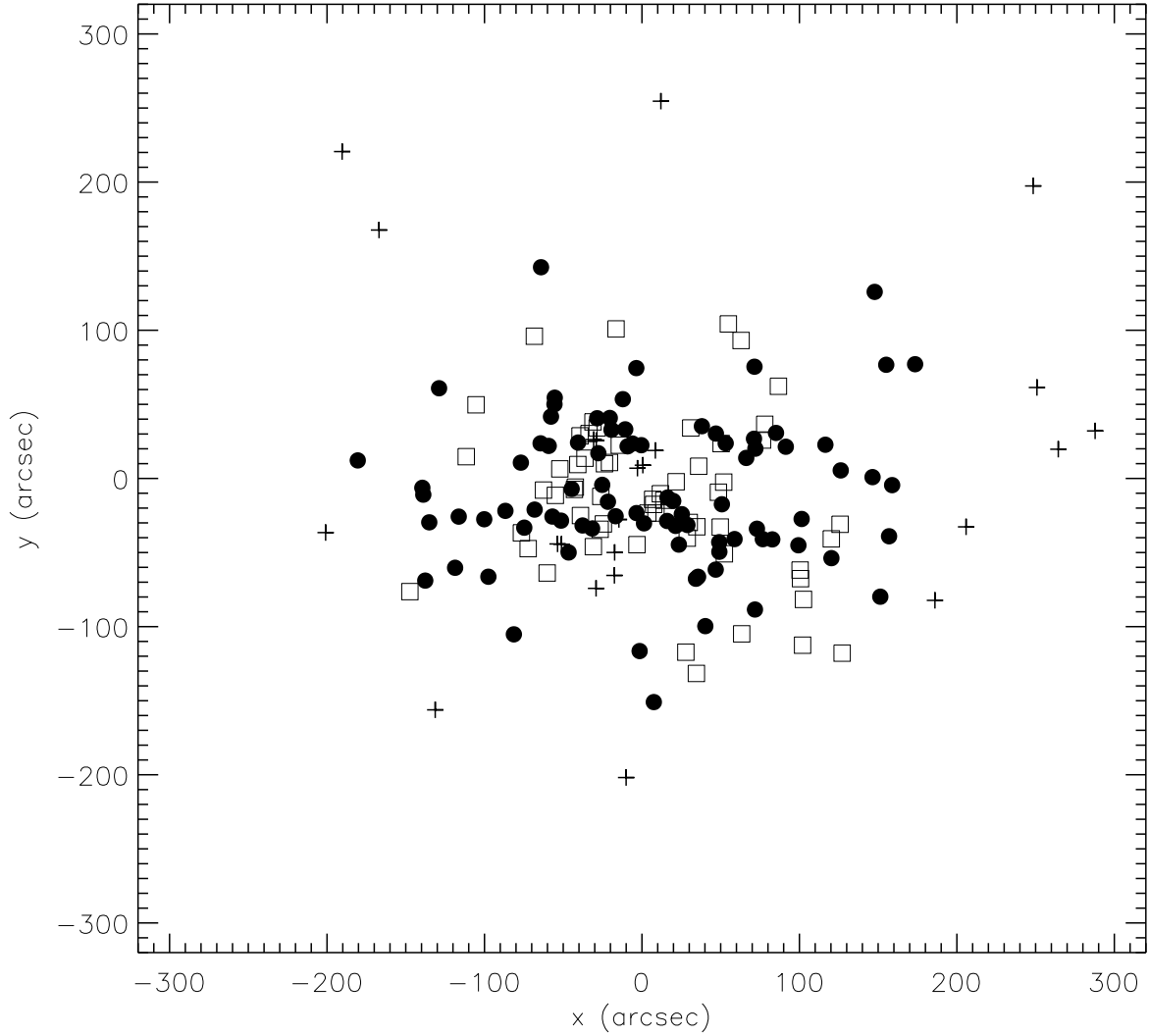}
\caption{Positions $x, y$ in arc seconds relative to the 
center of light of NGC 821, for the 168 PN candidates with 
measured velocities. The $x$-axis runs in the direction of increasing 
Right Ascension, along the major axis of NGC 821.
Plus signs represent PN.S detection only, open squares represent
FOCAS detection only, and filled circles are the PNs in common 
to both samples.
}
\end{figure}

\begin{figure}
\figurenum{8}
\epsscale{1.0}
\plotone{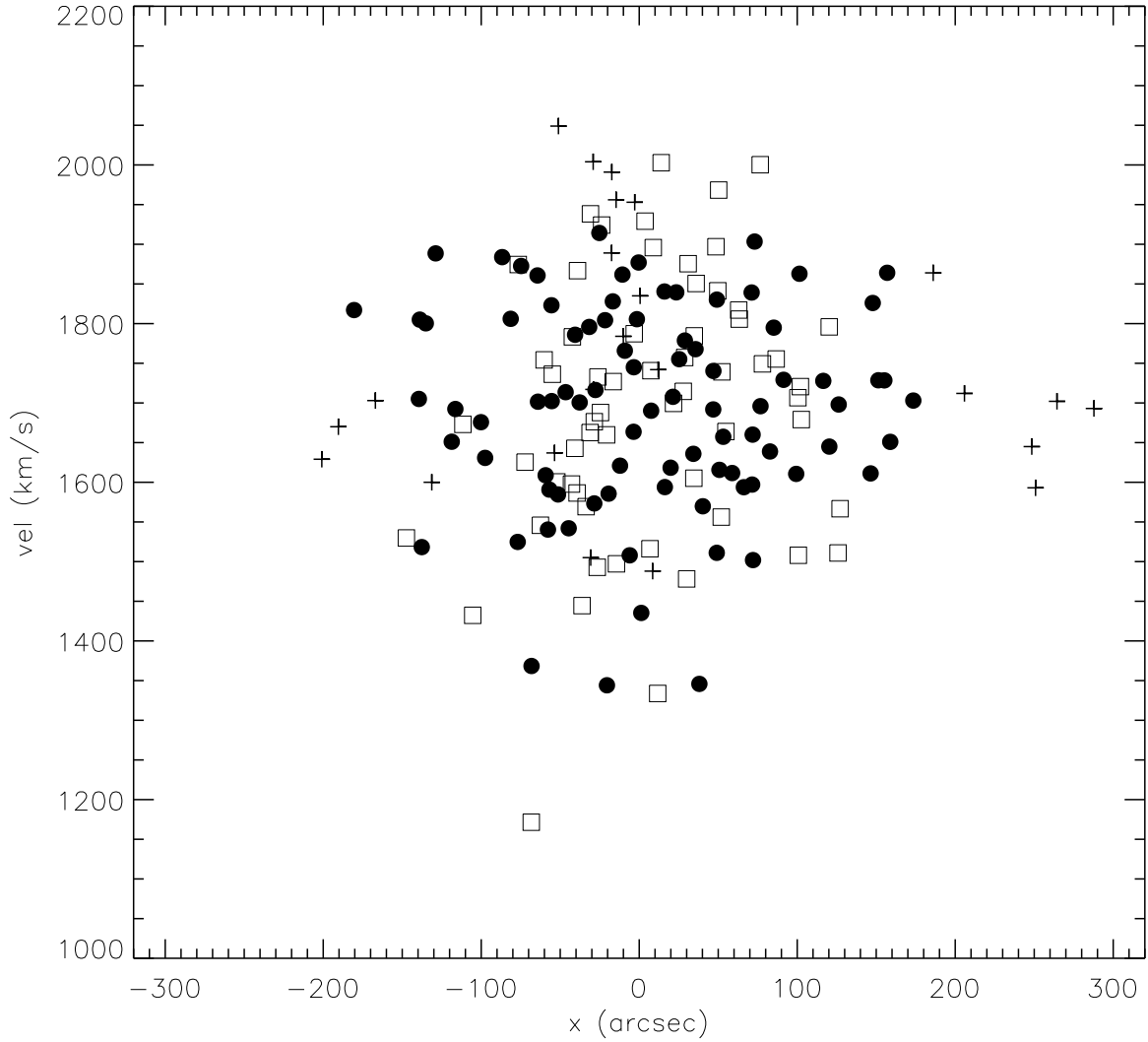}
\caption{Velocities of the 168 PN candidates as a function 
of their $x$-coordinates in arc seconds relative to the center of light 
of NGC 821. The symbols are the same as in figure 7. The isolated object
with the lowest velocity is FOCAS object 104 in Table 2.
}
\end{figure}

\begin{figure}
\figurenum{9}
\epsscale{1.0}
\plotone{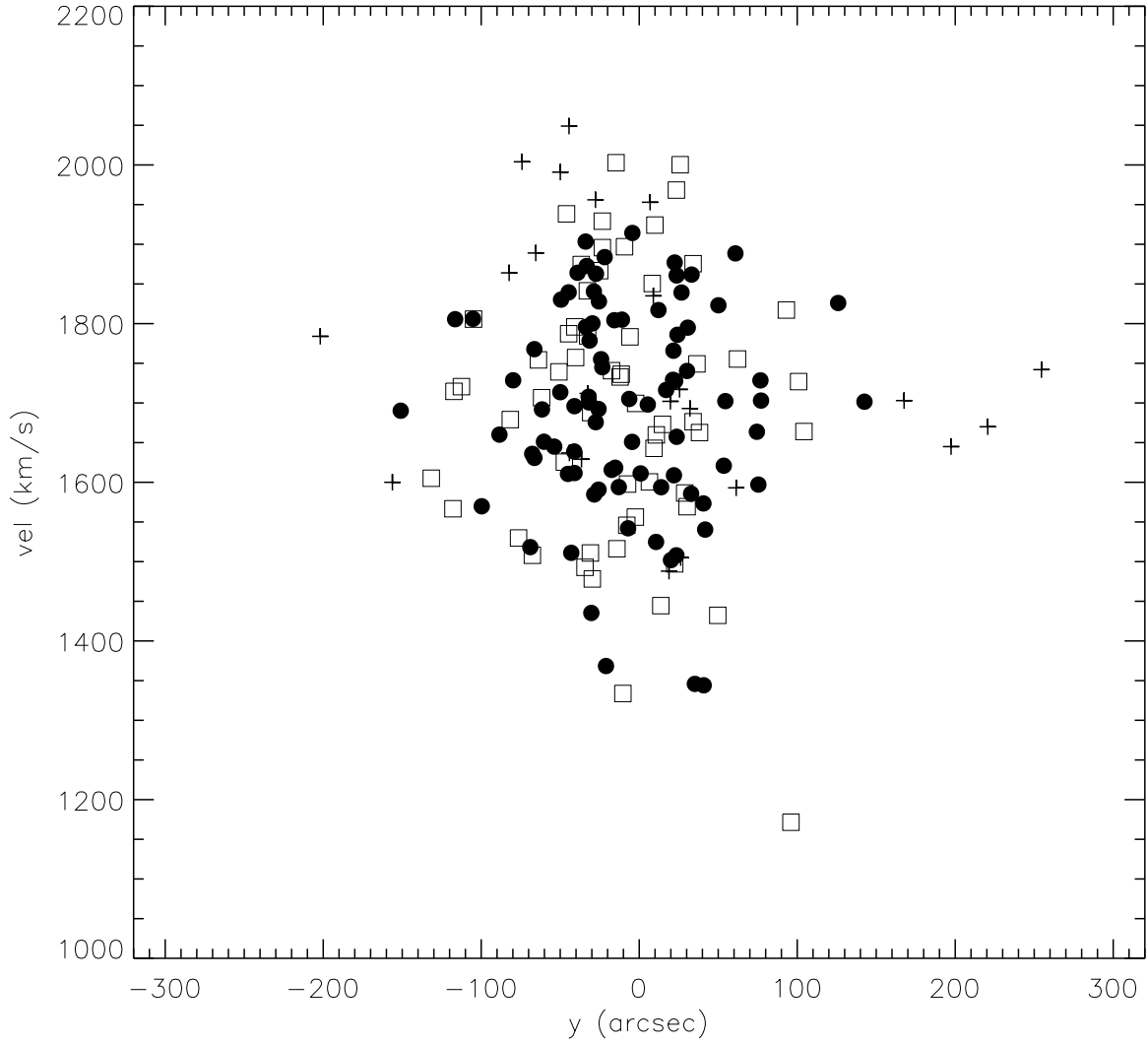}
\caption{Velocities of the 168 PN candidates as a function 
of their $y$-coordinates in arc seconds relative to the center of light 
of NGC 821. The symbols are the same as in figure 7. The isolated object
with the lowest velocity is FOCAS object 104 in Table 2.
}
\end{figure}

\begin{figure}
\figurenum{10}
\epsscale{1.0}
\plotone{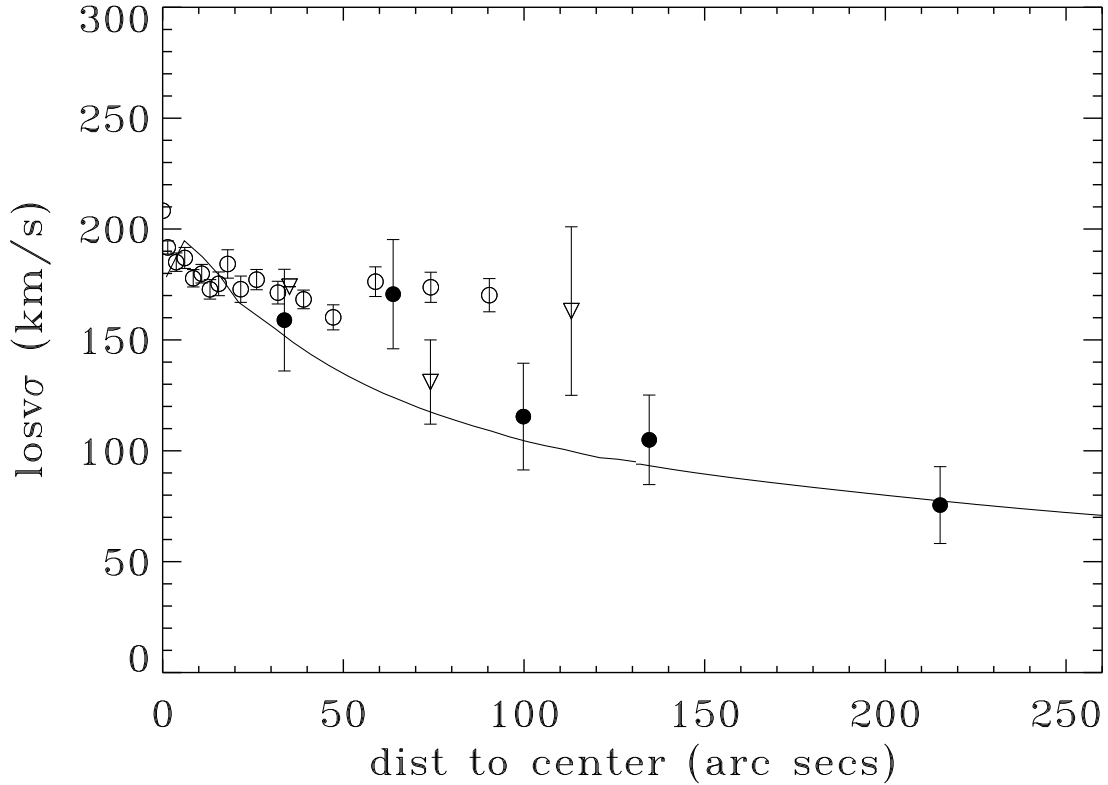}
\caption{Line-of-sight velocity dispersion as a function of 
projected angular distance to the center of NGC 821 for the FOCAS +
PN.S sample of 167 PNs (filled circles). 
The PNs were divided into 5 elliptical annuli as explained in the text.
Overplotted are Sauron data (triangles; Weijmans et al. 2009) and 
Forestell \& Gebhardt data (open circles). 
The solid line represents the analytical model of 
Hernquist (1990), with a constant $M/L$ ratio, a total mass of 
2 $\times$ 10$^{11}$ $M_{\odot}$, and $R_{\rm e}$= 39$''$.
}
\end{figure}

\begin{figure}
\figurenum{11}
\epsscale{1.0}
\plotone{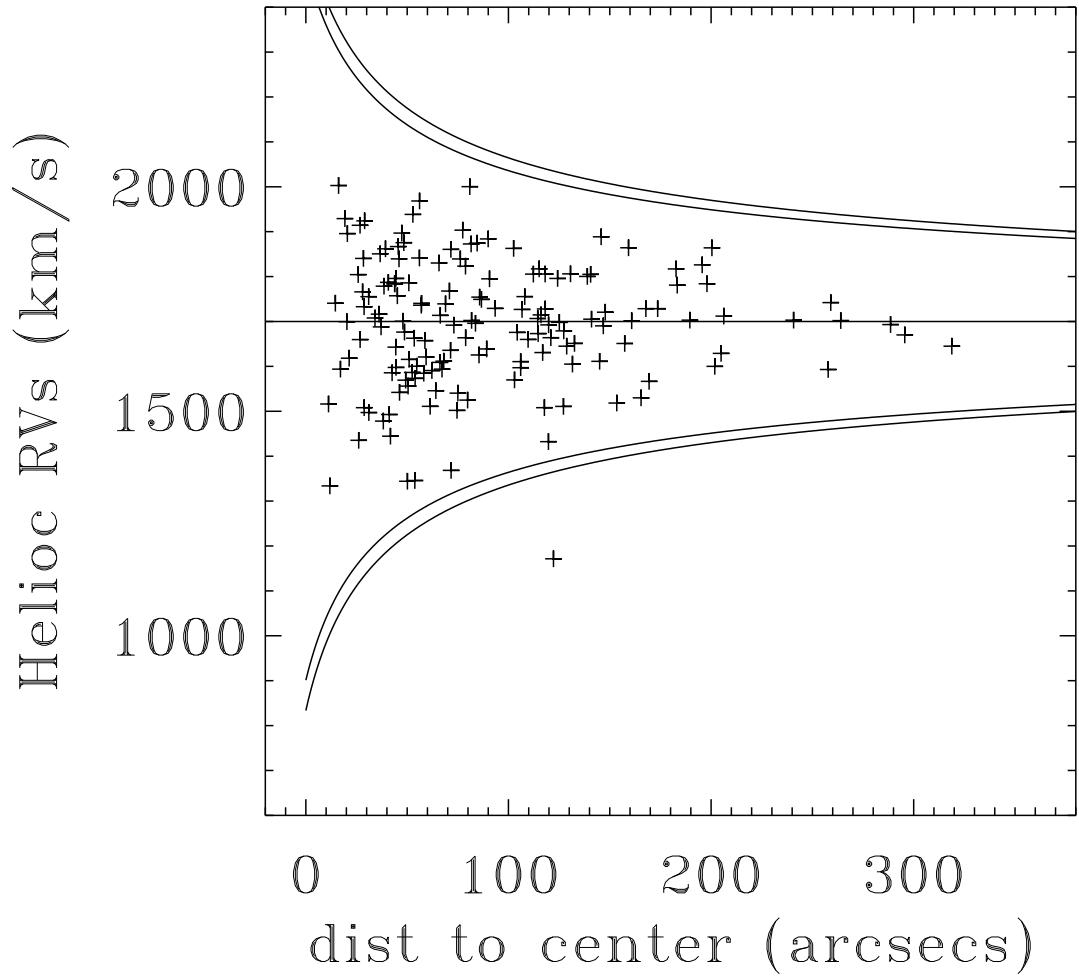}
\caption{Individual PN radial velocities plotted as a 
function of angular distance from the center of NGC 821. The solid 
lines are escape velocities for Hernquist models with total masses
2.0 x 10$^{11}$ $M_{\odot}$ (outer line) and 1.7 x 10$^{11}$ $M_{\odot}$ 
(inner line).
Note the outside position of the unrelated FOCAS object 104, which 
we have rejected as a PN. No other object has a velocity in excess
of the local escape velocity, even if we assume the lower mass.
}
\end{figure}

\begin{figure}
\figurenum{12}
\epsscale{1.0}
\plotone{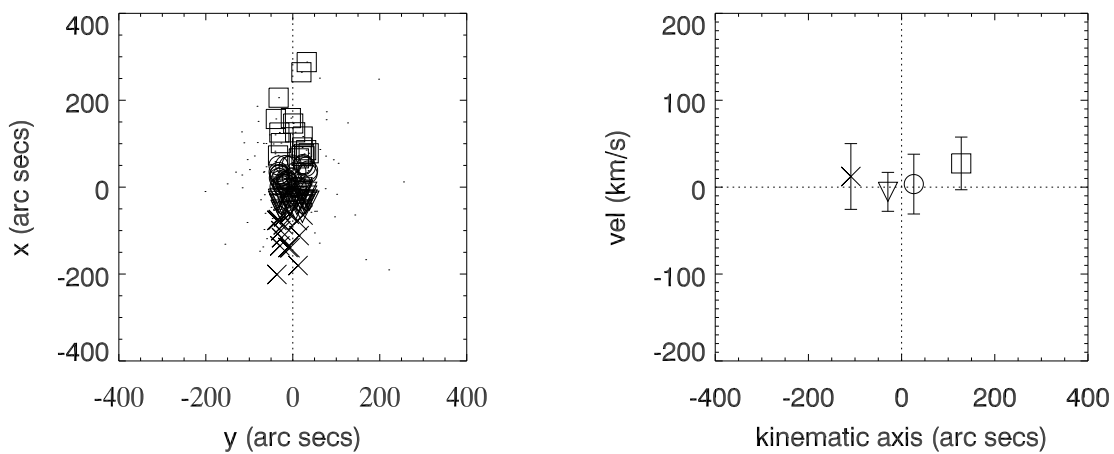}
\caption{Rotation of the PN.S + FOCAS sample of 167 PNs
along the photometric major axis. {\it Left}: distribution of PNs
in the sky, showing the same orientation as in Figure 3 of Coccato 
et al. (2009). The dotted line represents the major axis. We select all
PNs within $\pm 40''$ of the dotted line, and divide them in 4 bins.
Squares: $x > 60''$. Circles: $0'' < x < 60''$. 
Triangles: $-60'' < x < 0''$. Crosses: $x < -60''$.
{\it Right}: Average velocity for each of the four bins defined in the 
left figure. There is no significant rotation in this direction.
}
\end{figure}

\begin{figure}
\figurenum{13}
\epsscale{1.0}
\plotone{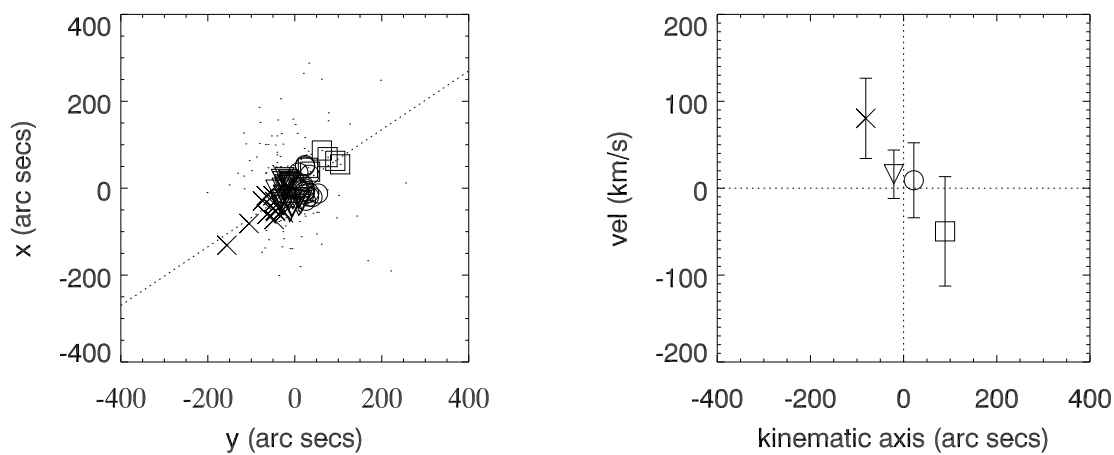}
\caption{Rotation of the PN.S + FOCAS sample of 167 PNs
along an axis inclined 56 degrees with respect to the photometric 
major axis. {\it Left}: distribution of PNs
in the sky, showing the same orientation as in Figure 3 of Coccato 
et al. (2009). The dotted line represents the inclined axis. We select all
PNs within $\pm 40''$ of the dotted line, calculate their $X$ coordinates 
with respect to the new axis, and divide them in 4 bins.
Squares: $X > 50''$. Circles: $0'' < X < 50''$. 
Triangles: $-50'' < X < 0''$. Crosses: $X < -50''$.
{\it Right}: Average velocity for each of the four bins defined in the 
left figure. The sense of rotation agrees with that reported by 
Coccato et al. (2009).
}
\end{figure}

\begin{figure}
\figurenum{14}
\epsscale{1.0}
\plotone{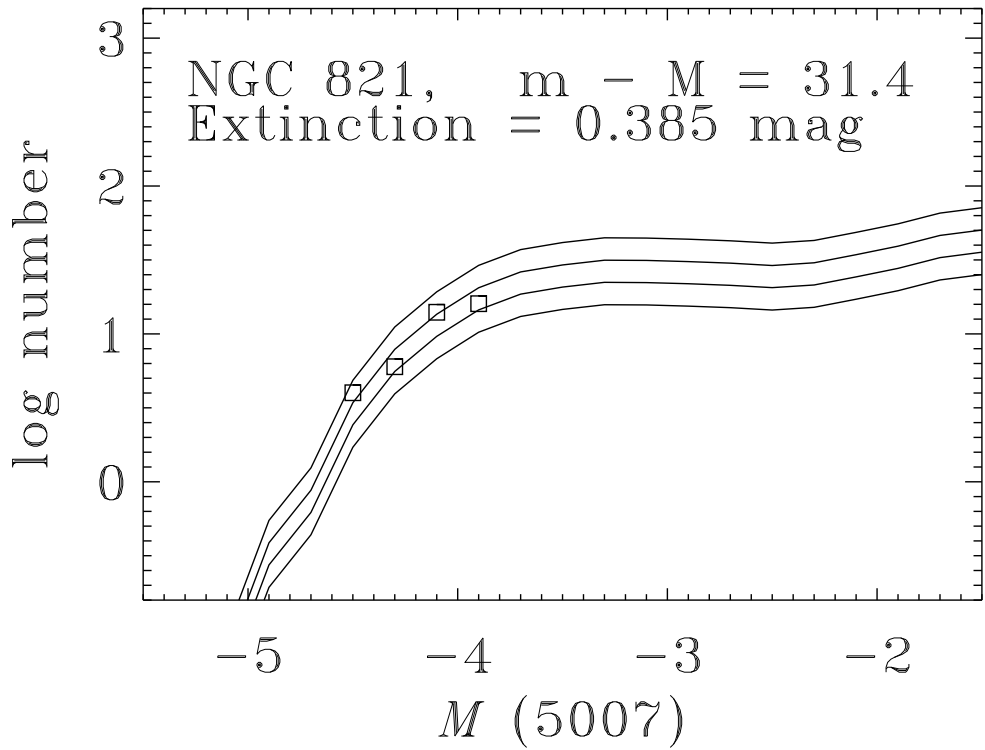}
\caption{Observed [O III] $\lambda$5007 PNLF of NGC 821 
(squares), with the statistically complete sample of 40 PNs binned 
into 0.2 mag intervals. The apparent magnitudes m(5007) have been 
transformed into absolute magnitudes M(5007) by adopting an extinction 
correction of 0.385 mag and a distance modulus m - M = 31.4. 
The 4 lines are PNLF simulations (M\'endez and Soffner 1997) for 
4 different total PN population sizes: 1200, 1700, 2400, and 3400 PNs. 
We estimate the best-fit sample size to be 2200. From this sample size 
it is possible to estimate the PN formation rate (see text).
}
\end{figure}

\begin{figure}
\figurenum{15}
\epsscale{1.0}
\plotone{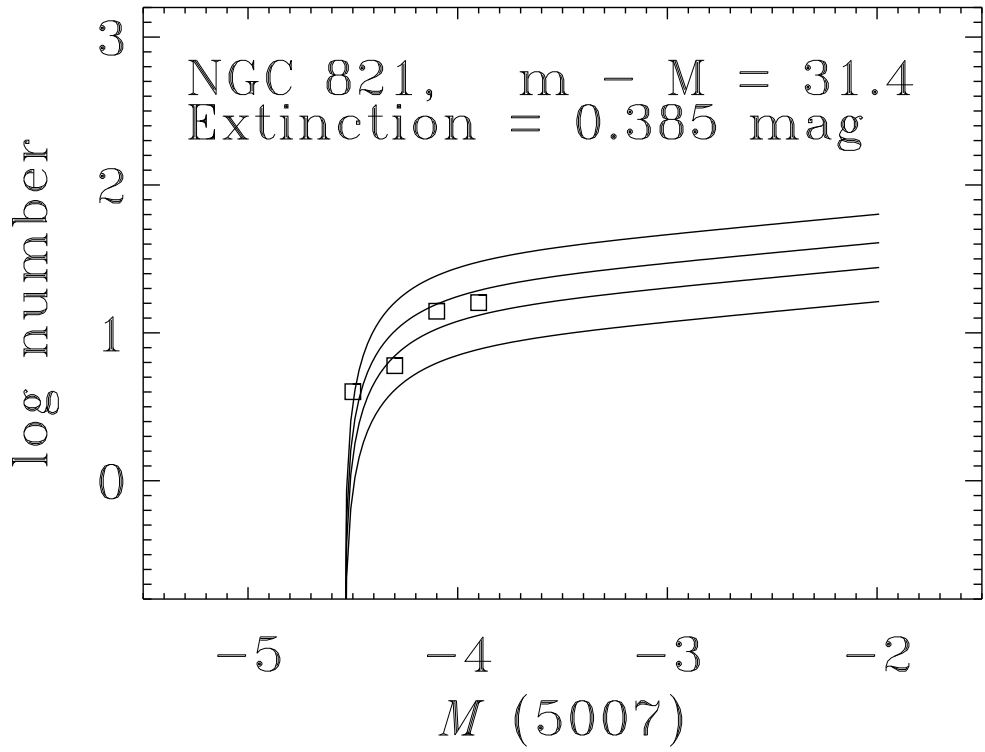}
\caption{Observed PNLF now compared with the analytical 
representation of the PNLF, using a universal cutoff at $-$4.5. 
The best-fit distance modulus is again 31.4.
}
\end{figure}

\begin{figure}
\figurenum{16}
\epsscale{1.0}
\plotone{f16.ps}
\caption{Difference between PNLF and SBF distance modulus
plotted as a function of SBF distance modulus, for 23 galaxies. 
Non-elliptical galaxies are plotted as diamonds, while for ellipticals 
we show just the error bars. The two galaxies near $m-M$=28 are NGC 5128 
and M 81. The only two galaxies with a positive $\Delta$mod are NGC 4258
and NGC 3115. The agglomeration of galaxies between $m-M$ 31 and 32 
is dominated by the Virgo and Fornax clusters. Our recent PNLF additions, 
ellipticals NGC 1344 in Fornax and NGC 821, are indicated as squares.
}
\end{figure}

%%%%%%%%%%%%%%%%%%%%%%%%%

\clearpage

\begin{deluxetable}{lrcrc}
\tablecaption{FOCAS observations and calibrations \label{tbl-1}}
\tablewidth{0pt}
\tablehead{
\colhead{FOCAS Field} & \colhead{Configuration} &
\colhead{FOCAS number} &
\colhead{exp (s)} & \colhead{Air mass\tablenotemark{a}}}
\startdata
 NGC 821            &  Off-band  &  55693  &  140  &  1.07 \\
 NGC 821            &   On-band  &  55695  & 1400  &  1.06 \\
 NGC 821            &  On+grism  &  55697  & 2100  &  1.03 \\
 NGC 821            &  Off-band  &  55705  &  140  &  1.01 \\
 NGC 821            &   On-band  &  55707  & 1400  &  1.01 \\
 NGC 821            &  On+grism  &  55709  & 2100  &  1.02 \\
 NGC 821            &  Off-band  &  55719  &  140  &  1.08 \\
 NGC 821            &   On-band  &  55721  & 1400  &  1.08 \\
 NGC 821            &  On+grism  &  55723  & 2100  &  1.13 \\
 NGC 821            &  Off-band  &  55729  &  140  &  1.28 \\
 NGC 821            &   On-band  &  55731  & 1400  &  1.30 \\
 NGC 821            &  On+grism  &  55733  & 2100  &  1.43 \\
 NGC 7293 + mask &   On-band  &  55855  &  200  &  1.36 \\
 NGC 7293 + mask &  On+grism  &  55857  &  300  &  1.35 \\
 NGC 7293 + mask &   On-band  &  55869  &  200  &  1.32 \\
 NGC 7293 + mask &  On+grism  &  55871  &  300  &  1.32 \\
 LTT 9491        &   On-band  &  55877  &   10  &  1.28 \\ 
 LTT 9491        &   On-band  &  55879  &   20  &  1.28 \\ 
 LTT 9491        &   On-band  &  55881  &   10  &  1.28 \\ 
 LTT 9491        &   On-band  &  55883  &   20  &  1.28 \\ 
 NGC 821            &  Off-band  &  55885  &  140  &  1.54 \\
 NGC 821            &   On-band  &  55887  & 1400  &  1.50 \\
 NGC 821            &  On+grism  &  55889  & 2100  &  1.35 \\
 NGC 821            &  Off-band  &  55897  &  140  &  1.17 \\
 NGC 821            &   On-band  &  55899  & 1400  &  1.16 \\
 NGC 821            &  On+grism  &  55901  & 2100  &  1.10 \\
 NGC 821            &  Off-band  &  55909  &  140  &  1.03 \\
 NGC 821            &   On-band  &  55911  & 1400  &  1.03 \\
 NGC 821            &  On+grism  &  55913  & 2100  &  1.02 \\
 NGC 821            &  Off-band  &  55919  &  140  &  1.02 \\
 NGC 821            &   On-band  &  55921  & 1400  &  1.02 \\
 NGC 821            &  On+grism  &  55923  & 2100  &  1.04 \\
 NGC 821            &  Off-band  &  55931  &  140  &  1.11 \\
 NGC 821            &   On-band  &  55933  & 1400  &  1.12 \\
 NGC 821            &  On+grism  &  55935  & 2100  &  1.19 \\
 NGC 821            &   On-band  &  55943  & 1400  &  1.38 \\
 NGC 821            &  On+grism  &  55945  & 2100  &  1.52 \\
NGC 7293 + mask &   On-band  &  56035  &  200  &  1.35 \\
NGC 7293 + mask &  On+grism  &  56037  &  300  &  1.34 \\
NGC 7293 + mask &   On-band  &  56043  &  200  &  1.33 \\
NGC 7293 + mask &  On+grism  &  56045  &  300  &  1.33 \\
 LTT 9491        &   On-band  &  56053  &   10  &  1.29 \\ 
 LTT 9491        &   On-band  &  56055  &   20  &  1.29 \\ 
 LTT 9491        &   On-band  &  56057  &   10  &  1.28 \\ 
 LTT 9491        &   On-band  &  56059  &   20  &  1.28 \\ 
NGC 821            &  Off-band  &  56061  &  140  &  1.55 \\
NGC 821            &   On-band  &  56063  & 1400  &  1.50 \\
NGC 821            &  On+grism  &  56065  & 2100  &  1.35 \\
NGC 821            &  Off-band  &  56073  &  140  &  1.18 \\
NGC 821            &   On-band  &  56075  & 1400  &  1.17 \\
NGC 821            &  On+grism  &  56077  & 2100  &  1.10 \\
NGC 821            &  Off-band  &  56085  &  140  &  1.04 \\
NGC 821            &   On-band  &  56087  & 1400  &  1.04 \\
NGC 821            &  On+grism  &  56089  & 2100  &  1.02 \\
NGC 821            &  Off-band  &  56097  &  140  &  1.02 \\
NGC 821            &   On-band  &  56099  & 1400  &  1.02 \\
NGC 821            &  On+grism  &  56101  & 2100  &  1.04 \\
NGC 821            &  Off-band  &  56111  &  140  &  1.10 \\
NGC 821            &   On-band  &  56113  & 1400  &  1.11 \\
NGC 821            &  On+grism  &  56115  & 2100  &  1.17 \\
NGC 821            &  Off-band  &  56123  &  140  &  1.35 \\
NGC 821            &   On-band  &  56125  & 1400  &  1.37 \\
NGC 821            &  On+grism  &  56127  & 2100  &  1.52 \\
\enddata
\tablenotetext{a}{the air masses correspond to the middle of each exposure}
\end{deluxetable}

\clearpage

%\begin{deluxetable}{rlllrllrrrc}
%\tablecaption{Detected Objects \label{tbl-2}}
%\tablewidth{0pt}
%\rotate
%\tabletypesize{\tiny}
%\tablehead{
%\colhead{Subaru id} &\colhead{PN.S id} &
%\colhead{\ } &  \colhead{$\alpha$} & \colhead{(2000)} & \colhead{\ } & 
%\colhead{$\delta$} &
%\colhead{(2000)} & \colhead{m(5007)} & \colhead{Helioc. RV} & {Notes}}

\begin{deluxetable}{c|c|rrr|rrr|c|cc}
\tablecaption{Detected PN Candidates \label{mytable}}
\tablewidth{0pt}
\rotate
\tabletypesize{\small}
\tablehead{

\colhead{ID} & \colhead{ID} &
\multicolumn{3}{c}{$\alpha$} & \multicolumn{3}{c}{$\delta$} & 
\colhead{$m$} & \colhead{Helioc. RV} & \colhead{Notes}       \\

\colhead{(FOCAS)} & \colhead{(PN.S)} &
\multicolumn{3}{c}{(J2000)} & \multicolumn{3}{c}{(J2000)} & 
\colhead{(5007) }& \colhead{(km s$^{-1}$)} & \colhead{}}

\startdata
101 &  ...&   2 &  8 &   7.78 &  10 &  59 &  41.53 & 28.46 &  ... &   \\
102 &  04 &   2 &  8 &  10.56 &  11 &   0 &   0.47 & 27.82 & 1701 &   \\
103 &  05 &   2 &  8 &  13.05 &  10 &  58 &  22.91 & 28.79 & 1888 &   \\
104 &  ...&   2 &  8 &  13.12 &  10 &  59 &  32.86 & 28.67 & 1171 & a \\
201 &  ...&   2 &  8 &  13.92 &  11 &   2 &  19.61 & 28.43 &  ... &   \\
105 &  06 &   2 &  8 &  14.08 &  10 &  57 &  13.57 & 27.78 & 1817 &   \\
106 & ... &   2 &  8 &  14.52 &  10 &  58 &  37.27 & 28.54 & 1432 &   \\
107 & ... &   2 &  8 &  14.65 &  11 &   0 &  19.84 & 27.44 & 1726 &   \\
108 &  07 &   2 &  8 &  15.99 &  10 &  59 &  22.59 & 27.74 & 1702 &   \\
109 & ... &   2 &  8 &  16.14 &  10 &  57 &  23.15 & 29.03 &  ... &   \\
110 &  08 &   2 &  8 &  16.24 &  10 &  59 &  20.18 & 27.49 & 1823 &   \\
111 & ... &   2 &  8 &  16.34 &  10 &  58 &  13.94 & 29.03 & 1672 &   \\
112 &  11 &   2 &  8 &  16.59 &  10 &  57 &  39.20 & 28.85 & 1705 &   \\
113 &  10 &   2 &  8 &  16.64 &  11 &   0 &  17.28 & 28.07 & 1663 &   \\
114 &  12 &   2 &  8 &  16.65 &  10 &  59 &  13.96 &  ...  & 1540 &   \\
115 &  13 &   2 &  8 &  16.88 &  10 &  57 &  37.33 & 28.42 & 1805 &   \\
202 & ... &   2 &  8 &  16.95 &  11 &   1 &  22.73 & 28.92 & 1663 &   \\
116 &  15 &   2 &  8 &  17.47 &  10 &  58 &  59.02 & 27.79 & 1860 &   \\
117 &  16 &   2 &  8 &  17.56 &  10 &  59 &  59.13 & 27.57 & 1620 &   \\
118 &  19 &   2 &  8 &  17.74 &  10 &  59 &  38.54 & 28.20 & 1573 &   \\
119 &  17 &   2 &  8 &  17.75 &  10 &  59 &   2.54 & 27.85 & 1608 &   \\
120 &  18 &   2 &  8 &  17.79 &  10 &  58 &  41.56 & 28.52 & 1524 &   \\
121 & ... &   2 &  8 &  17.79 &  10 &  59 &  35.02 & 28.07 & 1662 &   \\
122 & ... &   2 &  8 &  17.87 &  10 &  58 &   4.84 & 29.23 &  ... &   \\
203 & ... &   2 &  8 &  17.88 &  11 &   1 &  23.91 & 29.00 & 1817 &   \\
123 &  20 &   2 &  8 &  18.01 &  10 &  59 &  45.49 &  ...  & 1344 &   \\
124 & ... &   2 &  8 &  18.05 &  10 &  59 &  23.10 & 28.91 & 1586 &   \\
125 &  23 &   2 &  8 &  18.10 &  10 &  57 &  30.96 & 28.01 & 1800 &   \\
126 & ... &   2 &  8 &  18.12 &  10 &  59 &  35.16 & 27.92 & 1676 &   \\
127 & ... &   2 &  8 &  18.16 &  10 &  59 &  28.75 & 28.36 & 1569 &   \\
128 &  24 &   2 &  8 &  18.27 &  10 &  59 &  19.68 & 27.70 & 1785 &   \\
129 &  27 &   2 &  8 &  18.51 &  10 &  59 &  42.29 & 27.70 & 1585 &   \\
130 &  28 &   2 &  8 &  18.53 &  10 &  57 &  48.99 & 27.57 & 1692 &   \\
131 & ... &   2 &  8 &  18.65 &  10 &  58 &  37.13 & 29.06 &  ... &   \\
132 &  30 &   2 &  8 &  18.80 &  10 &  59 &  49.95 & 27.37 & 1861 &   \\
133 & ... &   2 &  8 &  18.90 &  10 &  59 &   0.64 &  ...  & 1600 &   \\
204 &  33 &   2 &  8 &  18.94 &  11 &   2 &  53.52 & 27.74 & 1826 & b \\
134 & ... &   2 &  8 &  19.04 &  10 &  59 &  17.99 &  ...  & 1444 &   \\
135 & ... &   2 &  8 &  19.13 &  10 &  59 &  11.90 & 27.86 & 1642 &   \\
136 &  34 &   2 &  8 &  19.14 &  10 &  59 &  27.02 & 27.05 & 1716 &   \\
137 &  35 &   2 &  8 &  19.20 &  10 &  58 &   1.96 & 28.37 & 1675 &   \\
205 &  36 &   2 &  8 &  19.21 &  11 &   1 &  22.08 & 28.22 & 1597 &   \\
138 & ... &   2 &  8 &  19.29 &  10 &  59 &  41.21 & 28.51 & 1497 &   \\
139 &  37 &   2 &  8 &  19.34 &  10 &  58 &  16.39 & 27.76 & 1883 &   \\
140 & ... &   2 &  8 &  19.37 &  10 &  58 &  44.33 & 28.75 & 1545 &   \\
141 & ... &   2 &  8 &  19.45 &  10 &  59 &  57.70 & 28.28 &  ... &   \\
142 &  39 &   2 &  8 &  19.52 &  10 &  59 &  48.95 & 28.22 & 1507 &   \\
143 &  38 &   2 &  8 &  19.52 &  10 &  59 &  45.35 & 27.98 & 1765 &   \\
144 & ... &   2 &  8 &  19.69 &  10 &  59 &  26.70 &  ...  & 1924 &   \\
145 & ... &   2 &  8 &  19.75 &  10 &  59 &  29.87 & 27.56 & 1659 &   \\
146 &  42 &   2 &  8 &  19.78 &  10 &  59 &  53.23 & 27.91 & 1877 & b \\
147 & ... &   2 &  8 &  19.84 &  10 &  58 &  48.97 & 28.66 & 1736 &   \\
148 &  44 &   2 &  8 &  19.94 &  10 &  58 &  32.74 & 27.79 & 1368 &   \\
149 &  43 &   2 &  8 &  19.95 &  10 &  59 &   0.10 & 27.54 & 1542 &   \\
150 & ... &   2 &  8 &  19.96 &  10 &  59 &   2.72 & 27.98 & 1783 &   \\
151 & ... &   2 &  8 &  20.03 &  10 &  59 &   1.43 &  ...  & 1597 &   \\
206 & ... &   2 &  8 &  20.20 &  11 &   0 &  26.10 & 28.02 & 1875 &   \\
152 &  47 &   2 &  8 &  20.30 &  10 &  57 &   8.53 & 28.45 & 1518 &   \\
207 &  46 &   2 &  8 &  20.38 &  11 &   0 &  32.76 & 27.35 & 1345 &   \\
153 & ... &   2 &  8 &  20.39 &  10 &  56 &  56.33 & 28.53 & 1529 &   \\
154 &  51 &   2 &  8 &  20.42 &  10 &  58 &  20.82 & 27.77 & 1872 &   \\
155 &  49 &   2 &  8 &  20.46 &  10 &  57 &  29.23 & 27.98 & 1651 &   \\
156 &  48 &   2 &  8 &  20.47 &  10 &  59 &  18.13 &  ...  & 1914 &   \\
208 & ... &   2 &  8 &  20.48 &  11 &   0 &   5.69 & 27.80 &  ... &   \\
209 & ... &   2 &  8 &  20.51 &  11 &   1 &  28.34 & 27.48 & 1755 &   \\
157 & ... &   2 &  8 &  20.56 &  10 &  58 &  17.47 & 28.58 & 1874 &   \\
158 &  52 &   2 &  8 &  20.61 &  10 &  58 &  40.01 & 27.36 & 1590 &   \\
159 & ... &   2 &  8 &  20.89 &  10 &  59 &  13.34 & 27.84 & 1732 &   \\
160 &  57 &   2 &  8 &  20.96 &  10 &  58 &  43.28 & 27.76 & 1584 &   \\
210 &  55 &   2 &  8 &  20.98 &  11 &   0 &  37.84 & 28.77 & 1740 &   \\
161 & ... &   2 &  8 &  21.19 &  10 &  58 &  55.63 &  ...  & 1866 &   \\
162 &  58 &   2 &  8 &  21.26 &  10 &  59 &  15.36 & 27.78 & 1804 &   \\
163 & ... &   2 &  8 &  21.33 &  10 &  58 &  15.64 & 28.43 & 1625 &   \\
164 & ... &   2 &  8 &  21.44 &  10 &  56 &  40.74 & 28.86 &  ... &   \\
211 & ... &   2 &  8 &  21.49 &  11 &   0 &  37.22 & 29.42 & 1968 &   \\
165 &  61 &   2 &  8 &  21.55 &  10 &  57 &  44.24 & 27.63 & 1630 &   \\
212 &  62 &   2 &  8 &  21.58 &  11 &   0 &  39.78 & 28.10 & 1657 &   \\
166 &  64 &   2 &  8 &  21.63 &  10 &  58 &  53.33 & 27.83 & 1700 &   \\
213 & ... &   2 &  8 &  21.70 &  11 &   1 &   7.64 & 28.33 & 1749 &   \\
214 & ... &   2 &  8 &  21.88 &  11 &   0 &  17.10 & 28.04 & 1850 &   \\
167 &  66 &   2 &  8 &  21.96 &  10 &  58 &  57.47 & 27.48 & 1795 &   \\
215 & ... &   2 &  8 &  21.99 &  10 &  59 &  59.42 & 27.88 & 1699 &   \\
168 &  65 &   2 &  8 &  22.00 &  10 &  59 &  14.53 & 28.54 & 1828 & b \\
169 & ... &   2 &  8 &  22.03 &  10 &  59 &   5.17 & 28.81 & 1687 &   \\
216 &  70 &   2 &  8 &  22.03 &  11 &   0 &  56.77 & 27.65 & 1839 &   \\
217 &  69 &   2 &  8 &  22.06 &  11 &   2 &  34.47 & 27.95 & 1728 &   \\
218 & ... &   2 &  8 &  22.11 &  10 &  59 &  46.72 & 27.73 & 1333 &   \\
219 & ... &   2 &  8 &  22.16 &  10 &  59 &  40.56 & 28.01 & 1516 &   \\
170 & ... &   2 &  8 &  22.17 &  10 &  59 &   1.54 & 28.40 & 1492 &   \\
220 & ... &   2 &  8 &  22.27 &  11 &   1 &   0.98 & 28.86 & 2000 &   \\
221 &  73 &   2 &  8 &  22.29 &  11 &   1 &  10.78 & 28.34 & 1794 &   \\
222 &  74 &   2 &  8 &  22.34 &  10 &  59 &  27.02 & 27.76 & 1745 & b \\
171 &  76 &   2 &  8 &  22.38 &  10 &  58 &  36.34 & 28.07 & 1713 &   \\
223 & ... &   2 &  8 &  22.38 &  10 &  59 &  39.30 & 28.59 & 1740 &   \\
224 &  78 &   2 &  8 &  22.42 &  10 &  59 &  49.24 & 27.66 & 1593 &   \\
225 & ... &   2 &  8 &  22.44 &  10 &  59 &  46.32 & 28.68 & 2002 &   \\
226 &  77 &   2 &  8 &  22.45 &  11 &   0 &  54.07 & 28.17 & 1501 &   \\
227 & ... &   2 &  8 &  22.52 &  10 &  59 &  29.94 & 28.45 &  ... &   \\
228 & ... &   2 &  8 &  22.59 &  10 &  59 &  33.11 & 28.84 & 1929 &   \\
229 &  79 &   2 &  8 &  22.61 &  11 &   0 &  45.90 & 28.52 & 1593 &   \\
230 &  81 &   2 &  8 &  22.68 &  10 &  59 &  51.18 & 28.63 & 1618 &   \\
231 &  80 &   2 &  8 &  22.68 &  11 &   2 &  50.39 & 27.72 & 1703 & b \\
172 & ... &   2 &  8 &  22.70 &  10 &  58 &  51.85 & 28.81 & 1938 &   \\
173 & ... &   2 &  8 &  22.71 &  10 &  58 &  17.55 & 28.29 & 1754 &   \\
232 & ... &   2 &  8 &  22.77 &  10 &  59 &  37.57 & 28.23 & 1895 &   \\
233 &  82 &   2 &  8 &  22.91 &  10 &  59 &  27.49 & 28.67 & 1435 &   \\
234 &  84 &   2 &  8 &  23.05 &  11 &   1 &  11.32 & 28.20 & 1729 &   \\
235 & ... &   2 &  8 &  23.06 &  11 &   0 &  25.24 & 28.54 & 1556 &   \\
236 &  85 &   2 &  8 &  23.33 &  10 &  59 &  40.92 &  ...  & 1840 &   \\
237 & ... &   2 &  8 &  23.34 &  11 &   0 &  18.83 & 28.22 & 1896 &   \\
238 &  86 &   2 &  8 &  23.39 &  10 &  59 &  51.25 & 27.80 & 1755 &   \\
239 & ... &   2 &  8 &  23.59 &  10 &  59 &  16.26 & 28.51 & 1787 &   \\
240 &  88 &   2 &  8 &  23.71 &  10 &  59 &  43.80 & 28.19 & 1707 &   \\
241 &  89 &   2 &  8 &  23.85 &  11 &   1 &  33.56 & 28.17 & 1727 &   \\
242 & ... &   2 &  8 &  23.88 &  10 &  59 &  52.44 & 28.91 & 1478 &   \\
243 &  90 &   2 &  8 &  23.89 &  11 &   0 &  16.52 & 28.67 & 1615 &   \\
244 &  91 &   2 &  8 &  23.94 &  10 &  59 &  50.57 & 28.19 & 1778 &   \\
245 & ... &   2 &  8 &  24.22 &  10 &  59 &  55.07 &  ...  & 1784 &   \\
174 &  93 &   2 &  8 &  24.38 &  10 &  57 &  38.09 & 27.97 & 1806 &   \\
246 & ... &   2 &  8 &  24.45 &  10 &  59 &  45.92 & 29.17 & 1757 &   \\
247 &  95 &   2 &  8 &  24.52 &  10 &  59 &  39.08 & 28.36 & 1839 &   \\
248 & ... &   2 &  8 &  24.75 &  11 &   0 &   7.70 & 28.70 & 1841 &   \\
249 &  98 &   2 &  8 &  25.20 &  11 &   1 &  32.99 & 28.39 & 1698 &   \\
250 &  99 &   2 &  8 &  25.32 &  11 &   0 &   1.91 & 27.84 & 1511 &   \\
251 & 101 &   2 &  8 &  25.54 &  11 &   0 &  11.31 & 27.92 & 1611 &   \\
252 & 102 &   2 &  8 &  25.63 &  11 &   0 &  27.11 & 28.46 & 1903 &   \\
253 & 104 &   2 &  8 &  25.70 &  10 &  59 &  58.63 & 28.14 & 1830 &   \\
254 & ... &   2 &  8 &  25.89 &  11 &   0 &   0.68 & 28.37 & 1739 &   \\
255 & 106 &   2 &  8 &  26.17 &  11 &   0 &  26.68 & 28.30 & 1695 &   \\
256 & 105 &   2 &  8 &  26.17 &  11 &   1 &  47.96 & 28.76 & 1611 &   \\
257 & 107 &   2 &  8 &  26.21 &  10 &  59 &  38.36 & 27.74 & 1767 &   \\
258 & 108 &   2 &  8 &  26.24 &  10 &  59 &  36.49 & 27.81 & 1635 &   \\
259 & 110 &   2 &  8 &  26.24 &  11 &   0 &  54.83 & 28.75 & 1862 &   \\
260 & 109 &   2 &  8 &  26.32 &  10 &  59 &  50.46 & 28.10 & 1691 &   \\
261 & 112 &   2 &  8 &  26.39 &  11 &   0 &  31.75 & 28.40 & 1638 &   \\
175 & ... &   2 &  8 &  26.62 &  10 &  57 &  53.60 & 28.75 &  ... &   \\
262 & 113 &   2 &  8 &  26.92 &  11 &   1 &  55.85 & 28.59 & 1650 &   \\
263 & 114 &   2 &  8 &  27.20 &  11 &   0 &  43.92 & 28.61 & 1610 &   \\
264 & ... &   2 &  8 &  27.30 &  11 &   1 &  13.91 & 27.90 & 1510 &   \\
265 & ... &   2 &  8 &  27.68 &  11 &   1 &   4.08 & 28.88 & 1795 &   \\
266 & 115 &   2 &  8 &  27.83 &  10 &  58 &  40.73 & 27.86 & 1805 &   \\
267 & ... &   2 &  8 &  28.21 &  11 &   0 &  36.22 & 28.64 & 1706 &   \\
268 & 116 &   2 &  8 &  28.31 &  10 &  59 &  25.15 & 27.86 & 1569 &   \\
269 & 117 &   2 &  8 &  28.44 &  11 &   0 &  57.42 & 27.64 & 1644 &   \\
270 & ... &   2 &  8 &  28.56 &  11 &   0 &  33.48 & 28.18 & 1508 &   \\
271 & 118 &   2 &  8 &  28.76 &  10 &  59 &  57.95 & 28.62 & 1660 &   \\
272 & 119 &   2 &  8 &  28.86 &  11 &   1 &  36.41 & 27.90 & 1864 &   \\
273 & ... &   2 &  8 &  28.90 &  10 &  59 &   5.50 & 27.34 & 1714 &   \\
274 & ... &   2 &  8 &  29.42 &  10 &  59 &  42.25 & 28.32 & 1805 &   \\
275 & ... &   2 &  8 &  29.44 &  11 &   0 &  27.86 & 28.77 & 1679 &   \\
276 & ... &   2 &  8 &  29.97 &  10 &  59 &   3.84 & 28.85 & 1605 &   \\
277 & 122 &   2 &  8 &  30.15 &  10 &  58 &  30.72 & 27.73 & 1690 &   \\
278 & 124 &   2 &  8 &  31.04 &  11 &   1 &  10.63 & 28.02 & 1728 &   \\
279 & ... &   2 &  8 &  31.22 &  11 &   0 &  11.48 & 28.95 & 1720 &   \\
280 & ... &   2 &  8 &  32.41 &  11 &   0 &  30.24 & 28.50 & 1566 &   \\ 
\enddata

Notes. (a) Rejected as PN: wrong velocity for NGC 821 if interpreted 
as 5007 \AA \ emitter. (b) Velocity taken from the PN.S study.
\end{deluxetable}


\begin{thebibliography}{}
\bibitem[1998]{alard} Alard, C., \& Lupton, R.H. 1998, ApJ, 503, 325

\bibitem[2001]{Bla01} Blakeslee, J.P., et al. 2001, MNRAS, 327, 1004

\bibitem[2009]{Bla09} Blakeslee, J.P., et al. 2009, ApJ, 694, 556 

\bibitem[2006]{buzzoni} Buzzoni, A., Arnaboldi, M., \& Corradi, R.L.M. 
         2006, MNRAS, 368, 877 

\bibitem[2002]{ciard1} Ciardullo, R., Feldmeier, J.J., Jacoby, G.H., et al.
         2002, ApJ, 577, 31

\bibitem[1999]{ciard2} Ciardullo, R., \& Jacoby, G.H. 1999, ApJ, 515, 191

\bibitem[2009]{coccato} Coccato, L., Gerhard, O., Arnaboldi, M., et al. 
         2009, MNRAS, 394, 1249 

\bibitem[1994]{colina} Colina, L., \& Bohlin, R.C. 1994, AJ, 108, 1931

\bibitem[2008]{delorenzi1} De Lorenzi, F., et al. 2008, MNRAS, 385, 1729

\bibitem[2009]{delorenzi2} De Lorenzi, F., et al. 2009, MNRAS, 395, 76

\bibitem[1991]{devaucouleurs} de Vaucouleurs, G., de Vaucouleurs, A., 
    Corwin, H.G., Jr., Buta, R.J., Paturel, G., \& Fouque, P. 1991, 
    Third Reference Catalogue of Bright Galaxies (Berlin: Springer) 

\bibitem[2005]{dekel} Dekel, A., et al. 2005, Nature, 437, 707

\bibitem[2004]{emsellem} Emsellem, E., Capellari, M., Peletier, R.F., 
     et al. 2004, MNRAS, 352, 721 

\bibitem[2007]{feldm} Feldmeier, J.J., Jacoby, G.H., \& Phillips, M.M.
         2007, ApJ, 657, 76

\bibitem[2008]{736} Forestell, A.D., \& Gebhardt, K. 2008, astro-ph 
                    arXiv:0803.3626

\bibitem[2003]{gebh} Gebhardt, K., et al. 2003, ApJ, 583, 92

\bibitem[2002]{gossl} G\"ossl, C.A., \& Riffeser, A.. 2002, A\&A, 381, 1095

\bibitem[1990]{hernquist} Hernquist, L. 1990, ApJ, 356, 359 

\bibitem[1995]{hui} Hui, X., Ford, H.C., Freeman, K.C., \& Dopita, M.A. 
         1995, ApJ, 449, 592 

\bibitem[1989]{jacoby} Jacoby, G.H. 1989, ApJ, 339, 39 

\bibitem[1990]{jacoby02} Jacoby, G.H., Ciardullo, R., \& Ford, H.C. 
         1990, ApJ, 356, 332 

\bibitem[1987]{jacoby01} Jacoby, G.H., Quigley, R.J., \& Africano, J.L. 
         1987, PASP, 99, 672 

\bibitem[2002]{742}
               Kashikawa, N., Aoki, K., Asai, R., et al. 2002, PASJ, 54, 819

\bibitem[2009]{kormen} Kormendy, J., Fisher, D.B., Cornell, M.E., \&
                       Bender, R. 2009, ApJS, 182, 216

\bibitem[2000]{kudri} Kudritzki, R.P., M\'endez, R.H, Feldmeier, J.J., et al.
         2000, ApJ, 536, 19

\bibitem[]{744}
         Meaburn, J., Boumis, P., L\'opez, J.A., et al. 2005, MNRAS, 360, 963

\bibitem[2001]{mendez02} M\'endez, R.H., Riffeser, A., Kudritzki, R.-P., 
                               et al. 2001, ApJ, 563, 135 

\bibitem[1997]{mendez01} M\'endez, R.H \& Soffner, T. 1997, A\&A, 321, 898 

\bibitem[2009]{mendez03} M\'endez, R.H, Teodorescu, A.M., Kudritzki, R.P., 
                         \& Burkert, A. 2009, ApJ, 691, 228

\bibitem[2005]{750} M\'endez, R.H., Thomas, D., Saglia, R.P., et al. 
         2005, ApJ, 627, 767

\bibitem[2003]{monet} Monet, D.J., et al. 2003, ApJ, 125, 984 

\bibitem[2005]{napo} Napolitano, N.R., et al. 2005, MNRAS, 357, 691

\bibitem[2009]{napo9} Napolitano, N.R., et al. 2009, MNRAS, 393, 329

\bibitem[1990]{oke} Oke, J.B. 1990, AJ, 99, 1621 

\bibitem[2004]{754} Peng, E.W., Ford, H.C., \& Freeman, K.C. 
         2004, ApJ, 602, 685

\bibitem[2009]{proctor} Proctor, R.N., et al. 2009, MNRAS, 398, 91

\bibitem[2003]{romanowsky} Romanowsky, A.J., Douglas, N.G., Arnaboldi, M., 
         et al. 2003, Science, 301, 1696 

\bibitem[1998]{schlegel} Schlegel, D.J., Finkbeiner, D.P., \& 
                         Davis, M. 1998, ApJ, 500, 525 

\bibitem[1987]{stetson} Stetson, P.B. 1987, PASP, 99, 191 

\bibitem[2005]{teodorescu} Teodorescu, A.M., M\'endez, R.H., Saglia, R.P.,
         et al. 2005, ApJ, 635, 290

\bibitem[2001]{tonry} Tonry, J.L., Dressler, A., Blakeslee, J.P., et al.
         2001, ApJ, 546, 681

\bibitem[2009]{weij} Weijmans, A.M., et al. 2009, MNRAS, 398, 561
\end{thebibliography}
\end{document}